\documentclass[12pt]{article}
\usepackage{epsfig,cite}
\usepackage{amsmath}
\usepackage{amssymb}
\usepackage{color}
\usepackage{graphicx}
\usepackage{verbatim,mathrsfs,subfigure,feynmf}
\usepackage{url}
\usepackage{ytableau}
\include{epsf}
\newcount\timecount
\newcount\hours \newcount\minutes  \newcount\temp \newcount\pmhours
\hours = \time
\divide\hours by 60
\temp = \hours
\multiply\temp by 60
\minutes = \time
\advance\minutes by -\temp
\def\hour{\the\hours}
\def\minute{\ifnum\minutes<10 0\the\minutes
            \else\the\minutes\fi}
\def\clock{
\ifnum\hours=0 12:\minute\ AM
\else\ifnum\hours<12 \hour:\minute\ AM
      \else\ifnum\hours=12 12:\minute\ PM
            \else\ifnum\hours>12
                 \pmhours=\hours
                 \advance\pmhours by -12
                 \the\pmhours:\minute\ PM
                 \fi
            \fi
      \fi
\fi
}

\def\monthname{\relax\ifcase\month 0/\or January\or February\or
   March\or April\or May\or June\or July\or August\or September\or
   October\or November\or December\else\number\month/\fi}

\def\bold#1{\setbox0=\hbox{$#1$}%
     \kern-.025em\copy0\kern-\wd0
     \kern.05em\copy0\kern-\wd0
     \kern-.025em\raise.0433em\box0 }

\def\beq{\begin{equation}}
\def\eeq{\end{equation}}

\def\ga{\mathrel{\raise.3ex\hbox{$>$\kern-.75em\lower1ex\hbox{$\sim$}}}}
\def\la{\mathrel{\raise.3ex\hbox{$<$\kern-.75em\lower1ex\hbox{$\sim$}}}}
\def\gev{{\rm \, Ge\kern-0.125em V}}
\def\tev{{\rm \, Te\kern-0.125em V}}
\def\gyr{{\rm \, G\kern-0.125em yr}}

\def\gappeq{\mathrel{\rlap {\raise.5ex\hbox{$>$}}
{\lower.5ex\hbox{$\sim$}}}}
\def\lappeq{\mathrel{\rlap{\raise.5ex\hbox{$<$}}
{\lower.5ex\hbox{$\sim$}}}}
\def\Toprel#1\over#2{\mathrel{\mathop{#2}\limits^{#1}}}

\def\m12{m_{1\!/2}}

\def\bea{\begin{eqnarray}}
\def\eea{\end{eqnarray}}

\usepackage{lscape}

\def\beq{\begin{equation}}
\def\eeq{\end{equation}}

\begin{document}\begin{titlepage}
\pagestyle{empty}
\begin{flushright}
{\tt KCL-PH-TH/2021-76}, {\tt CERN-TH-2021-154}  \\
{\tt ACT-4-21, MI-HET-765} \\
{\tt UMN-TH-4105/21, FTPI-MINN-21/22} \\
\end{flushright}

\begin{center}{\bf \large{Flipped SU(5) GUT Phenomenology: Proton Decay and $\mathbf{g_\mu - 2}$}}\\
\vskip 0.2in
{\bf John~Ellis}$^{a}$,
{\bf Jason~L.~Evans}$^{b}$,
{\bf Natsumi Nagata}$^{c}$, \\
{\bf Dimitri~V.~Nanopoulos}$^{d}$ and
{\bf Keith~A.~Olive}$^{e}$
\vskip 0.2in
{\small
{\em $^a$Theoretical Particle Physics and Cosmology Group, Department of
  Physics, King's~College~London, London WC2R 2LS, United Kingdom;\\
Theoretical Physics Department, CERN, CH-1211 Geneva 23,
  Switzerland;\\
National Institute of Chemical Physics and Biophysics, R\"{a}vala 10, 10143 Tallinn, Estonia}\\[0.2cm]
  {\em $^b$Tsung-Dao Lee Institute, Shanghai Jiao Tong University, Shanghai 200240, China}\\[0.2cm] 
  {\em $^c$Department of Physics, University of Tokyo, Bunkyo-ku, Tokyo
 113--0033, Japan}\\[0.2cm] 
{\em $^d$George P. and Cynthia W. Mitchell Institute for Fundamental
 Physics and Astronomy, Texas A\&M University, College Station, TX
 77843, USA;\\ 
 Astroparticle Physics Group, Houston Advanced Research Center (HARC),
 \\ Mitchell Campus, Woodlands, TX 77381, USA;\\ 
Academy of Athens, Division of Natural Sciences,
Athens 10679, Greece}\\[0.2cm] 
{\em $^e$William I. Fine Theoretical Physics Institute, School of
 Physics and Astronomy,\\ University of Minnesota, Minneapolis, MN 55455,
 USA}}
 
\vspace{0.5cm}
{\bf Abstract}
\end{center}
\baselineskip=18pt \noindent
{\small
We consider proton decay and $g_\mu - 2$ in flipped SU(5) GUT models.
We first study scenarios in which the soft supersymmetry-breaking
parameters are constrained to be universal at some high scale
$M_{in}$ above the standard GUT scale where the QCD and electroweak
SU(2) couplings unify. In this case the proton lifetime is typically
$\gtrsim 10^{36}$~yrs, too long to be detected in the foreseeable
future, and the supersymmetric contribution to $g_\mu - 2$ is too
small to contribute significantly to resolving the discrepancy between
the experimental measurement and data-driven calculations within the
Standard Model. However, we identify a region of the constrained
flipped SU(5) parameter space with large couplings between the 10- and
5-dimensional GUT Higgs representations where $p \to e^+ \pi^0$ decay
may be detectable in the Hyper-Kamiokande experiment now under construction,
though the contribution to $g_\mu -2$ is still small. A substantial
contribution to $g_\mu - 2$ is possible, however, if the universality
constraints on the soft supersymmetry-breaking masses are relaxed.
We find a `quadrifecta' region where observable proton decay
co-exists with a (partial) supersymmetric resolution
of the $g_\mu - 2$ discrepancy and acceptable values
of $m_h$ and the relic LSP density.}


\vfill
\leftline{October 2021}
\end{titlepage}

\section{Introduction}
\label{sec:intro}

The flipped SU(5) Grand Unified Theory (GUT) was first proposed in~\cite{Barr}, as a possible
intermediate gauge group obtained from the breaking of an underlying SO(10) GUT group. Flipped
SU(5) was subsequently investigated in~\cite{DKN} as a GUT group in its own right, independently
of a possible SO(10) parent group. Gauge coupling unification and the predictions for $\sin^2 \theta_W$
in flipped SU(5) models with and without supersymmetry were also studied in~\cite{DKN}. The
supersymmetric version of flipped SU(5) was subsequently advocated in~\cite{flipped2} on several grounds.
One was that breaking the initial GUT symmetry down to the Standard Model (SM) SU(3)$\times$SU(2)$\times$U(1)
gauge group via $\mathbf{10}$ and $\overline{\mathbf{10}}$ Higgs representations led to suppression of
proton decay via dimension-5 operators thanks to an economical missing-partner mechanism. It was
also argued that flipped SU(5) would fit naturally into string theory, as weakly-coupled string
models could not accommodate the adjoint and larger Higgs representations required to break other
GUT groups such as SU(5), SO(10) and E$_6$, but could accommodate the $\mathbf{10}$ and $\overline{\mathbf{10}}$
of flipped SU(5). Indeed, variants of flipped SU(5) were subsequently derived in the fermionic
formulation of weakly-coupled heterotic string theory~\cite{AEHN,recent}.

Following the formulation of flipped SU(5) and its derivation from string
theory, there have been many phenomenological studies of the model.
These have included such particle-physics topics as proton decay~\cite{Ellis:2020qad,Mehmood:2020irm,Haba:2021rzs}
and neutrino masses and mixing~\cite{Ellis:1993ks,egnno2,egnno3,egnno4}, as well as
cosmological issues such as dark matter~\cite{ehkno,emo3}, baryogenesis, inflation and
entropy generation~\cite{Campbell:1987eb,egnno2,egnno3,egnno4,FSU5Cosmo}. The upshot of these studies is that flipped SU(5)
can provide a complete framework for particle physics and cosmology
below the Planck scale. An additional topic of current interest in flipped SU(5) is
the muon anomalous magnetic moment, $g_\mu - 2$. It has been shown recently that the
discrepancy between the experimental measurement~\cite{BNL,FNAL} and the data-driven theoretical
calculation within the SM~\cite{datadriven} can be partially resolved within a minimal flipped SU(5)
model~\cite{eenno1} (where many relevant references can be found), and that the discrepancy with the experimental
measurement is completely resolved if the lattice SM 
calculation~\cite{BMW} is adopted.
The discrepancy is also completely resolved within non-minimal flipped 
SU(5) models~\cite{extend,recent} even if the data-driven SM
calculation~\cite{datadriven} is adopted. 

Motivated by this encouraging backdrop, in this paper we pursue
further studies of proton decay in supersymmetric flipped SU(5), which we link to an
investigation of the muon anomalous magnetic moment, $g_\mu - 2$. It was pointed out in the initial paper on the 
non-supersymmetric version of flipped SU(5)~\cite{Barr} that it predicted the same proton
decay modes as conventional SU(5), but with characteristic differences in the branching fractions
(see also~\cite{Ellis:2020qad}).
As already mentioned, dimension-5 contributions to proton decay are suppressed in supersymmetric
flipped SU(5), so the dimension-6 modes such as $p \to e^+ \pi^0$ are expected to dominate
proton decays in this model. A new generation of underground detectors with increased sensitivities
to this and other proton decay modes are now under construction, led by
Hyper-Kamiokande~\cite{HK}, so it is interesting to evaluate
accurately the expected rates for $p \to e^+ \pi^0$ and other proton decays. In this paper we
address two important aspects of such calculations, namely the appropriate matching conditions
at the GUT scale (see~\cite{emo3} for an earlier study), and the uncertainties associated with
SM input parameters and calculations of hadronic matrix elements~\cite{AISS}, which had been
examined previously in the context of conventional SU(5) in~\cite{Ellis:2019fwf}.

The outline of this paper is as follows. In Section~\ref{sec:model} we recall briefly the salient
features of the minimal supersymmetric flipped SU(5) model. Then, in Section~\ref{sec:match}
we discuss the GUT-scale matching conditions for the gauge couplings, Yukawa couplings and
soft supersymmetry-breaking parameters of the model, assuming that these are initially specified
at some input scale, $M_{in}$, above the scale where the SU(3) and SU(2) couplings of the SM
are unified. Section~\ref{sec:pdk} presents the formulae for the expressions relevant to the
calculations of the proton decay rates, including their uncertainties, and Section~\ref{sec:results} presents our
results. 

In the first version of the model that we study,
in Section~\ref{sec:universal}, the values of the soft
supersymmetry-breaking parameters
used as inputs at the input scale $M_{in} > M_{GUT}$ are
constrained to be universal \cite{super-GUT,emo,emo3,Ellis:2016qra,Ellis:2016tjc,Ellis:2017djk,Ellis:2019fwf}. In this case we find that
the proton lifetime is generally beyond the reach of the
next generation of experiments. However, the decay
$p \to e^+ \pi^0$ may be accessible if the couplings
$\lambda_{4,5}$ between the GUT Higgs fields in the
$\mathbf{10}$ and $\overline{\mathbf{10}}$ representations
and the SM Higgs fields in the $\mathbf{5}$ and
$\overline{\mathbf{5}}$ representations are both
relatively large, $\lambda_{4,5} \sim 0.5$. However,
even in this case the supersymmetric contribution to
$g_\mu -2$ is far smaller than the discrepancy between
the experimental value and that from data-driven or lattice
theoretical calculations in the SM \cite{eenno1}. We therefore
discuss in Section~\ref{sec:nonuniversal} the possibilities for the combination of detectable proton
decay and a substantial contribution to $g_\mu -2$
in flipped SU(5) with non-universal input soft
supersymmetry-breaking parameters. We find that the
$p \to e^+ \pi^0$ decay rate is quite insensitive to
the degree of non-universality, whereas this can allow
a much larger contribution to $g_\mu -2$, as illustrated
previously in~\cite{eenno1}. We exhibit `quadrifecta' domains of the
multi-dimensional unconstrained flipped SU(5)
parameter space where observable proton decay can
co-exist with a (partial) supersymmetric resolution
of the $g_\mu - 2$ discrepancy, while the calculated value
of $m_h$ is compatible with experiment within conservative calculational
uncertainties and the relic LSP density is similar to the
observed value.

Finally, Section~\ref{sec:conx} summarizes our conclusions.

\section{The Model}
\label{sec:model}

The model we consider is the minimal supersymmetric flipped SU(5)(FSU(5)) GUT
with the gauge symmetry SU(5)$\times$U(1)$_X$ \cite{Barr,DKN,flipped2,AEHN,recent,emo3, egnno2, egnno3, egnno4,eenno1},
where U(1)$_X$ is an `external' Abelian gauge factor.  Here, we review only the essential components of the model. The model contains three generations of minimal supersymmetric Standard Model (MSSM) matter fields, together with three right-handed neutrino chiral superfields. These are embedded into $\mathbf{10}$, $\bar{\mathbf{5}}$ and $\mathbf{1}$ representations, which are denoted by $F_i$, $\bar{f}_i$, and $\ell^c_i$, respectively, with $i=1,2,3$ the generation index. 
The SU(5) and U(1)$_X$ charges of the matter sector of the theory are
\begin{eqnarray}
\bar f_i (\bar{\bf 5},-3)&=&\left\{ U_i^c, L_j \left(U_{l}\right)_{ji}\right\} \nonumber \\ F_i({\bf 10},1)& = &\left\{ Q_i, V^{CKM}_{ij} e^{-i \varphi_j} D_j^c, \left(U_{\nu^c}\right)_{ij} N_j^c\right\} \, , \nonumber \\ 
   l^c_i({\bf 1},5) &=& \left(U_{l^c}\right)_{ij} E_j^c \, .
   \label{assign}
\end{eqnarray} 
A characteristic feature of the FSU(5) GUT is that the assignments of the quantum numbers for right-handed leptons and the right-handed up- and down-type quarks are ``flipped'' with respect to their assignments in standard SU(5). 
In Eq.~(\ref{assign}),
 the $V^{CKM}_{ij}$ are the Cabibbo-Kobayashi-Maskawa (CKM) matrix elements,
$U_{\nu^c}$, $U_l$, and $U_{l^c}$ are unitary matrices, and the phase
factors $\varphi_i$ satisfy the condition $\sum_{i} \varphi_i = 0$~\cite{Ellis:1993ks}. The
components of the doublet fields $Q_i$ and $L_i$ are written as 
\begin{equation}
 Q_i = 
\begin{pmatrix}
 u_i \\ V_{ij} d_j
\end{pmatrix}
~, \qquad
L_i = 
\begin{pmatrix}
 (U_{\rm PMNS})_{ij} \nu_j \\ e_i
\end{pmatrix}
~,
\end{equation}
where $U_{\rm PMNS}$ is the Pontecorvo-Maki-Nakagawa-Sakata (PMNS)
matrix.~\footnote{We define the PMNS matrix as in
the Review of Particle Physics (RPP)~\cite{PDG}, and note that $U_{\rm PMNS} = U^*_{\rm MNS}$ in the notation of Ref.~\cite{Ellis:1993ks}.}

The FSU(5) theory must be broken to the SM gauge symmetry.  This is accomplished by including
 a pair of $\mathbf{10}$ and $\overline{\mathbf{10}}$ Higgs fields, $H$ and $\overline{H}$, respectively, with
 the decompositions
 \begin{eqnarray}
H({\bf 10},1)=\left\{Q_H,D^c_H,N_H^c\right\} \, , \quad \quad \bar H(\overline {\bf 10},-1)=\left\{\bar Q_H,\bar D^c_H,\bar N_H^c\right\}~.
\end{eqnarray}
We note that the phase transition associated with this symmetry breaking
was discussed in detail in \cite{Campbell:1987eb,egnno2,egnno3}.
We recall also that the supersymmetric SM Higgs bosons are embedded in  a pair of $\mathbf{5}$ and $\overline{\mathbf{5}}$ Higgs multiplets, $h$ and $\bar{h}$, respectively, with the decompositions
\begin{eqnarray}
h({\bf 5},-2)= \left\{{H_c},H_d\right\} \, , \quad \quad \bar h(\overline{\bf 5},2)=\left\{ {\bar H_c}, H_u\right\} \, .
\end{eqnarray}
In addition, the theory has three (or more) SU(5) singlets $\phi_a$ that generate the masses of the right-handed neutrinos.
 
%
The superpotential for this theory is
\begin{align} \notag
W &=  \lambda_1^{ij} F_iF_jh + \lambda_2^{ij} F_i\bar{f}_j\bar{h} +
 \lambda_3^{ij}\bar{f}_i\ell^c_j h + \lambda_4 HHh + \lambda_5
 \bar{H}\bar{H}\bar{h}\\ 
&\quad   + \lambda_6^{ia} F_i\bar{H}\phi_a + \lambda_7^a h\bar{h}\phi_a
 + \lambda_8^{abc}\phi_a\phi_b\phi_c + \mu_\phi^{ab}\phi_a\phi_b\,, 
\label{Wgen} 
\end{align}
where the indices $i,j$ run over the three fermion families, the indices $a, b, c$
have ranges $\ge 3$, and for
simplicity we have suppressed gauge group indices.
We note that we have imposed
a $\mathbb{Z}_2$ symmetry  
$H\rightarrow -H$
to prevent the Higgs colour triplets or elements of the Higgs decuplets from mixing with SM fields. This symmetry also suppresses the
supersymmetric mass term for $H$ and $\overline{H}$, and thus suppresses dimension-five proton decay operators.
The first three terms of the superpotential (\ref{Wgen}) provide the SM Yukawa couplings, and the
fourth and fifth terms in (\ref{Wgen}) account for the splitting of the triplet and doublet masses in the Higgs 5-plets. The masses of the color triplets are
\begin{eqnarray}
M_{H_C}=4\lambda_4V \quad \quad \quad M_{\bar H_C}=4\lambda_5 V \, ,
\end{eqnarray}
where $V$ is the common vacuum expectation value (vev) of the $H$ and $\overline{H}$ fields that break FSU(5), 
with $V=\langle N^c_{H}\rangle =\langle {\bar N}^c_{H}\rangle$. 
The sixth term in (\ref{Wgen}) accounts for neutrino masses, and the seventh term plays the role of the 
$\mu$-term of the MSSM. The last two terms may play roles in cosmological inflation, along with $\lambda_6$, and 
also play roles in neutrino masses.
GUT symmetry breaking, inflation, leptogenesis, and the generation of neutrino masses in this model have been discussed 
recently in~\cite{egnno2,egnno3,egnno4,FSU5Cosmo}, and are reviewed in~\cite{building}.

\section{Matching Conditions}
\label{sec:match}

As a preliminary to giving the gauge coupling matching conditions, we first specify the masses of the fields that get masses from the breaking of the unified gauge symmetries. After the symmetry breaks, just as in the minimal SU(5) case, the heavy $X,\overline X$ gauge bosons of the SU(5) symmetry will mediate proton decay. We recall that the conventional SM hypercharge is a linear combination of the U(1)$_X$ gauge symmetry and the diagonal U(1) subgroup of SU(5):
\begin{eqnarray}
\frac{Y}{2}= \frac{1}{\sqrt{15}}Y_{24} +\sqrt{\frac{8}{5}} Q_X \, ,
\end{eqnarray}
where the $Q_X$ charge is in units of $\frac{1}{\sqrt{40}}$ and
\begin{eqnarray}
Y_{24}=\sqrt{\frac{3}{5}} {\rm diag}\left(\frac{1}{3},\frac{1}{3},\frac{1}{3},-\frac{1}{2},-\frac{1}{2}\right)~.
\end{eqnarray}
The gauge bosons that acquire masses from the breaking of SU(5)$\times$U(1)$\to$ SU(3)$\times$SU(2)$\times$U(1) are $X(3,2)_{1/3},\bar X(\bar 3,2)_{-1/3}$ and a singlet $V_1$ with masses
\begin{eqnarray}
M_X=g_5V\quad \quad \quad M_{V_1}=\sqrt{\frac{5}{2}}\left(\frac{24}{25}g_5^2+\frac{1}{25}g_X^2\right)^{1/2} V \, ,
\label{eq:MX0}
\end{eqnarray}
where $g_5$ and $g_X$ are the SU(5) and U(1)$_X$ gauge coupling constants, respectively, and $V$ is the (common)
vev of the $\mathbf{10}$ and $\overline{\mathbf{10}}$ Higgs fields. 

The gauge coupling matching conditions are 
\begin{eqnarray}
&&\frac{1}{g_1^2}=\frac{1}{25}\frac{1}{g_5^2}+\frac{24}{25}\frac{1}{g_X^2} +\frac{1}{8\pi^2}\left(\frac{4}{5}\ln\left[\frac{M_{GUT}}{\sqrt{M_{H_C}M_{\bar H_C}}}\right]-\frac{2}{5} \ln\left[\frac{M_{GUT}}{M_X}\right]\right) \, , \label{eq:g1}\\
&&\frac{1}{g_2^2}=\frac{1}{g_5^2}-\frac{6}{8\pi^2}\ln\left[\frac{M_{GUT}}{M_X}\right] \, , \label{eq:g2}\\
&& \frac{1}{g_3^2}=\frac{1}{g_5^2}+\frac{1}{8\pi^2}\left(2\ln\left[\frac{M_{GUT}}{\sqrt{M_{H_C}M_{\bar H_C}}}\right]-4\ln\left[\frac{M_{GUT}}{M_X}\right]\right) \, , \label{eq:g3}
\end{eqnarray}
with $M_{GUT}$ taken to be the renormalization scale where $g_2=g_3$.
Using this scale simplifies the analysis of the gauge matching conditions.  
Combining Eq. (\ref{eq:g2}) and (\ref{eq:g3}), we find
\begin{eqnarray}
M_X=\frac{\mu^2}{\sqrt{M_{H_C}M_{\bar H_C}}}\exp\left[4\pi^2\left(\frac{1}{g_2^2}-\frac{1}{g_3^2}\right)\right] \, , \label{eq:MX}
\end{eqnarray}
where $\mu$ is a renormalization scale that we can choose equal to $M_{GUT}$, in which case the 
matching at the scale $g_2(M_{GUT})=g_3(M_{GUT})$ would cause the exponent to disappear, yielding the rather simple relationship
$M_X=\mu^2/\sqrt{M_{H_C}M_{\bar H_C}}$. In our numerical calculations we match at a scale close to $M_{GUT}$, in which case the exponent has a small effect on this relationship. Keeping the exponential correction, we use the expression in Eq. (\ref{eq:g2}) and (\ref{eq:MX}) to obtain 
\begin{eqnarray}
\frac{1}{g_5^2}+\frac{3}{8\pi^2}\ln(g_5)-\frac{3}{2}\frac{1}{g_3^2}+\frac{1}{2}\frac{1}{g_2^2}-\frac{3}{8\pi^2}\ln(4\sqrt{\lambda_4\lambda_5})=0 \, .
\end{eqnarray}
We then solve this equation numerically for $g_5$, which can then be used in Eq. (\ref{eq:MX}) to obtain the vev:
\begin{eqnarray}
V=\frac{\mu}{2(\lambda_4\lambda_5)^{1/4}g_5^{1/2}}\exp\left[2\pi^2\left(\frac{1}{g_2^2}-\frac{1}{g_3^2}\right)\right] \, .
\end{eqnarray}
Once we have the vev and $g_5$, we can obtain $M_X$ from Eq.~(\ref{eq:MX0}).
In general, the loop corrections used in the matching conditions are important when the scale $M_{in}$ at which
universality is imposed on the soft supersymmetry-breaking parameters is $> M_{GUT}$. 

The SM Yukawa couplings are also matched at $M_{GUT}$, to $\lambda_{1,2,3}$ \cite{emo3,eenno1}:
\bea
 h_t = h_{\nu} = \lambda_2 /\sqrt{2} ~,\qquad 
 h_b = 4 \lambda_1 ~, \qquad 
 h_\tau = \lambda_3 ~.
\label{matching1}
\eea
Unlike minimal SU(5), the neutrino Yukawa couplings are naturally fixed 
to be equal to the up-quark
Yukawa couplings.~\footnote{See~\cite{egnno2,egnno3,FSU5Cosmo} for a discussion of inflation, supercosmology and neutrino masses in no-scale FSU(5).}
This is a consequence of the flipping that puts the right-handed neutrinos  
into decuplets in FSU(5), instead of being singlets as
in minimal SU(5), where their Yukawa couplings would be viewed as independent parameters. 

The supersymmetric FSU(5) GUT model is specified by
the following GUT-scale parameters.
There are two independent soft supersymmetry-breaking gaugino masses, namely a common mass $M_5$ 
for the SU(5) gauginos $\tilde g, \tilde W$ and $\tilde B$, and another mass $M_{X1}$
for the `external' gaugino $\tilde B_X$ that is independent {\it a priori}. There are also three independent soft
supersymmetry-breaking scalar masses that we assume to be generation-independent, 
namely $m_{10}$ for the sfermions in the $\mathbf{10}$ 
representations of SU(5), $m_{\bar 5}$ for the sfermions in the $\mathbf{\overline 5}$
representations of SU(5), and $m_1$ for the right-handed sleptons in the SU(5)-singlet
representations. We assume for simplicity that the trilinear soft supersymmetry-breaking parameters $A_0$ are universal.

We assume initially that the gaugino and scalar masses and trilinear parameters are
universal at $M_{in}$ \cite{emo3,FSU5Cosmo}, i.e., we take 
$M_5 = M_{X1} = m_{1/2}$ and $m_{10} = m_{\bar 5} = m_1 = m_0$ at 
$M_{in}$. 
In general, as in the NUHM2~\cite{nuhm2}, one may assume independent soft supersymmetry-breaking for the $\mathbf{5}$
and $\mathbf{\overline 5}$ Higgs representations, $m_{H_{1,2}}$. However, we begin by assuming that these are also universal so that $m_{H_{1}}=m_{H_{2}} = m_{10} = m_{\bar 5} = m_1$. We treat the ratio of SM Higgs vevs,
$\tan \beta$, as a free parameter. Finally, we assume that the Higgs mixing parameter $\mu > 0$,
so as to obtain a supersymmetric contribution to $g_\mu - 2$ with the same sign as the discrepancy between the
experimental measurement and the data-driven theoretical value in the SM.

The matching conditions for the the soft supersymmetry-breaking gaugino mass terms at $M_{GUT}$ are 
\begin{align}
 M_1 &= \frac{1}{25}\frac{g_1^2}{g_5^2} M_5+\frac{24}{25}\frac{g_1^2}{g_X^2} M_{X1}
-\frac{g_1^2}{16\pi^2}\left[\frac{2}{5} M_5 
 -\frac{2}{5}\left(A_4+A_5\right)\right] \, ,
\label{eq:m1match}
\\[3pt]
M_2 &= \frac{g_2^2}{g_5^2} M_5
-\frac{g_2^2}{16\pi^2}\left[6 M_5 \right] ~,
\label{eq:m2match}
\\[3pt]
M_3 &= \frac{g_3^2}{g_5^2} M_5
-\frac{g_3^2}{16\pi^2}\left[4 M_5 -\left(A_4+A_5\right) \right]
~,
\label{eq:m3match}
\end{align}
where $A_4$ and $A_5$ are the trilinear $A$-terms associated with the superpotential couplings $\lambda_4$ and $\lambda_5$, respectively.

We note that there are additional 1-loop contributions to the gaugino masses that could in principle be as large as the those included in Eqs.~(\ref{eq:m1match}-\ref{eq:m3match}).
These are proportional to the soft mass term along the 
flat direction ($\Phi = \langle N^c_H \rangle = \langle {\bar N}^c_H \rangle$) that breaks the FSU(5) gauge symmetry.  For example, $M_2$ includes an additional term, 
$+(g_2^2/16\pi^2)[6 m_\Phi/\sqrt{14}]$ on the right-hand side of Eq.~(\ref{eq:m2match}), and there are similar contributions for $M_{1,3}$.
As described in detail in \cite{egnno2,egnno3}, this flat direction is lifted by a non-renormalizable superpotential term
of the form $\lambda (H {\bar H})^n/M_P^{2n-3}$ with $n\ge4$ to obtain a sufficiently large vev, where $M_P$ is the reduced Planck mass, $M_P^2 = 1/8\pi G_N$. We expect
the soft mass for $\Phi$ to be of the same order as the other soft mass parameters, in which case the late decay of the flat direction releases entropy leading to a dilution factor of order $\Delta = 10^4 (m_\Phi/10~{\rm TeV})$, and a temperature (after decay) of about 1 MeV $\lambda_i^2$ ($m_\Phi/10~{\rm TeV})^{1/2}$. Lowering $m_\Phi$ would require 
some Yukawa coupling (e.g., $\lambda_7$) to be increased to maintain a temperature $\gtrsim$ 1 MeV, and would decrease $\Delta$
and the contribution to the gaugino masses.  However, due to the model dependence of $m_\Phi$, we do not include this contribution in Eqs.~(\ref{eq:m1match}-\ref{eq:m3match}).

The scalar soft masses are matched using \cite{emo3,eenno1}:
\begin{align}
     m_{Q}^2=m_{D}^2=m_{N}^2 &= m_{10}^2 ~, \quad 
m_{U}^2=m_{L}^2 = m_{5}^2 \, , \quad  
 m_{E}^2 = m_{1}^2~,   \nonumber \\
 m_{H_u}^2 &= m_{h_2}^2~, \quad 
 m_{H_d}^2 = m_{h_1}^2 \, .
\label{matching2}
\end{align}
The trilinear terms are initially set to be universal
at $M_{in}$ with $A_i = A_0$, corresponding to the Yukawa couplings $\lambda_i$ for $i = 1-5$. Each $A_i$ is run down to the GUT scale and matched using
\beq
 A_t=A_\nu = A_2, \qquad A_b = A_1, \qquad  A_\tau = A_3 \, .
\label{matching3}
\eeq
Finally, the magnitude of the MSSM $\mu$-term and the bilinear soft supersymmetry-breaking $B$-term
are determined at the electroweak scale by the
minimization of the Higgs potential. This also determines the pseudoscalar Higgs mass, $M_A$,
which we use as an input in {\tt FeynHiggs~2.18.10}~\cite{FH}
to determine the masses of the remaining physical Higgs degrees of freedom.~\footnote{Equivalently,
as in~\cite{nuhm2}, one can treat $\mu$ and $M_A$
as input parameters and use the minimization conditions to solve for the two Higgs soft masses.}

Our constrained FSU(5) model is therefore specified by the following set of parameters:
\beq
m_{1/2}, \, m_{0}, \, A_0, \, \tan\beta, \, M_{in}, \, \lambda_4, \, \lambda_5, \, \lambda_6 \, .
\label{CFSU5params}
\eeq
Later, in Section~\ref{sec:nonuniversal}, we generalize the model 
to allow $M_5 \ne M_{X1}$, as well as allowing 
the soft masses $m_{H_{1}}, m_{H_{2}},  m_{10}, m_{\bar 5}$, and  $m_1$ to differ from each other. 
The relevant RGEs for flipped SU(5) were given in \cite{emo3}. In principle,
it is also necessary to specify the mass of the heaviest left-handed neutrino,
$m_{\nu_3}$, which we take to be 0.05~eV. This and $\lambda_6$ fix the right-handed neutrino mass 
and $\mu_\phi$. However, our results
are quite insensitive to this choice.

\section{Proton Decay and Error Estimates}
\label{sec:pdk}

\subsection{Proton Lifetime}
\label{sec:tau}

Proton decay in FSU(5) was discussed in detail in  \cite{Ellis:2020qad}, and we quote
here only the essential results from that work.
Thanks to the suppression of dimension-5 operators by the FSU(5) missing-partner mechanism,
the main contribution to nucleon decay is due to the exchanges of SU(5) 
gauge bosons.~\footnote{The contribution of the color-triplet Higgs multiplets to the dimension-6 operators is negligible unless their masses are $\lesssim {\cal O}(10^{13})$~GeV~\cite{Hamaguchi:2020tet}.   } 
The relevant gauge interaction terms are 
\begin{align}
 K_{\rm gauge}&=
\sqrt{2}g_5\bigl(
-\epsilon_{\alpha\beta}(U^c_a)^{\dagger}X^\alpha_a U_l^T L^\beta
+\epsilon^{abc}(Q^{a\alpha})^{\dagger} X^\alpha_bV^{CKM} P^\dagger {D}^c_c
+ \epsilon_{\alpha\beta}(N^c)^\dagger U_{\nu^c}^\dagger X^\alpha_aQ^{a\beta}
+{\rm h.c.}
\bigr)~,
\label{eq:gaugeintflipped}
\end{align}
where 
the $X_a^\alpha$ are the
SU(5) gauge vector superfields,  
$P_{ij} \equiv e^{i\varphi_i} \delta_{ij}$, 
$\alpha, \beta$ are SU(2)$_L$ indices, and $a,b,c$ are
SU(3)$_C$ indices.
The relevant effective operator in FSU(5) below the electroweak scale is~\footnote{The operator $(u_L d_L) (u_R l_{Ri}) $ is not induced in FSU(5). } 
\begin{align}
 {\cal L}(p\to \pi^0 l^+_i)
&=C_{RL}(udul_i)\bigl[\epsilon_{abc}(u_R^ad_R^b)(u_L^cl_{Li}^{})\bigr]
~,
\label{eq:lagptopil}
\end{align}
where the Wilson coefficient can be written as
\beq
C_{RL}(udu\ell_i) =  \frac{g_5^2}{M_X^2} (U_l)_{i1} V^{CKM*}_{11}
 e^{i\varphi_1} ~,
\eeq
evaluated at the weak scale.

The partial proton decay widths to $\ell_i^+ \pi^0$ can be expressed as follows in terms of these coefficients at the hadronic scale:
\begin{equation}
 \Gamma (p\to  \ell^+_i \pi^0)=
\frac{m_p}{32\pi}\biggl(1-\frac{m_\pi^2}{m_p^2}\biggr)^2
\vert {\cal A}_L(p\to \ell^+_i \pi^0) \vert^2
~,
\label{eq:aptopil}
\end{equation}
where
\begin{align}
 {\cal A}_L(p\to \ell^+_i \pi^0)&=
C_{RL}(udu\ell_i)\langle \pi^0\vert (ud)_Ru_L\vert p\rangle
~,
\end{align}
and we use the following determinations of the matrix elements by
lattice calculations \cite{AISS}:
\begin{equation}
    \langle \pi^0|(ud)_Ru_L|p\rangle_e = -0.131(4)(13), \; \; \langle \pi^0|(ud)_Ru_L|p\rangle_\mu = -0.118(3)(12) \, .
\end{equation}
For proton decays with a final-state lepton $\ell_i$ ($\ell_1 = e, \, \ell_2 = \mu$), we have
\begin{align}
  \Gamma (p\to  \ell_i^+ \pi^0)_{\rm flipped}&=
\frac{g_5^4m_p |V_{ud}|^2 |(U_l)_{i1}|^2
}{32\pi M_X^4}\biggl(1-\frac{m_\pi^2}{m_p^2}\biggr)^2
A_L^2 A_{S_1}^2 \left(\langle \pi^0\vert (ud)_Ru_L\vert
 p\rangle_{\ell_i}\right)^2~,
\label{eq:ptopiefl}
\end{align}
where $m_p$ and $m_\pi$ denote the proton and pion masses,
respectively, and the subscript on the hadronic matrix element indicates
that it is evaluated at the corresponding lepton kinematic point. 
The renormalization factor 
between the GUT scale and the electroweak scale
is \cite{Munoz:1986kq, Abbott:1980zj}
\begin{align}
 A_{S_1} &=
 \biggl[
\frac{\alpha_3(\mu_{\text{SUSY}})}{\alpha_3(\mu_{\rm GUT})}
\biggr]^{\frac{4}{9}}
\biggl[
\frac{\alpha_2(\mu_{\text{SUSY}})}{\alpha_2(\mu_{\rm GUT})}
\biggr]^{-\frac{3}{2}}
\biggl[
\frac{\alpha_1(\mu_{\text{SUSY}})}{\alpha_1(\mu_{\rm GUT})}
\biggr]^{-\frac{1}{18}}
\nonumber \\
&\times
\biggl[
\frac{\alpha_3(m_Z)}{\alpha_3(\mu_{\rm SUSY})}
\biggr]^{\frac{2}{7}}
\biggl[
\frac{\alpha_2(m_Z)}{\alpha_2(\mu_{\rm SUSY})}
\biggr]^{\frac{27}{38}}
\biggl[
\frac{\alpha_1(m_Z)}{\alpha_1(\mu_{\rm SUSY})}
\biggr]^{-\frac{11}{82}} ~,
\end{align}
where $m_Z$, $\mu_{\rm SUSY}$, and $\mu_{\rm GUT}$ denote the $Z$-boson
mass, the SUSY scale and the GUT scale, respectively, and $\alpha_A
\equiv g_A^2/(4\pi)$ with $g_A$ ($A = 1,2,3$) the gauge coupling
constants of the SM gauge groups.  
Below the electroweak
scale, we take into account the perturbative QCD renormalization
factor, which was computed in Ref.~\cite{Nihei:1994tx} at the two-loop
level to be $A_L = 1.247$. 

Using Eq.~\eqref{eq:ptopiefl}, we can readily compute the partial lifetime of 
the $p\to e^+ \pi^0$ mode as \cite{Ellis:2020qad}:
\begin{align}
    \tau (p\to e^+ \pi^0)_{\rm flipped}&\simeq 
    7.9 \times 10^{35} \times  |(U_l)_{11}|^{-2}
    \biggl(\frac{M_X}{10^{16}~{\rm GeV}}\biggr)^4
     \biggl(\frac{0.0378}{\alpha_5}\biggr)^2 ~{\rm yrs}~.
     \label{tp0ef}
\end{align}
A similar expression can be obtained for the partial lifetime of 
the $p\to \mu^+ \pi^0$ mode:
\begin{align}
    \tau (p\to  \mu^+ \pi^0)_{\rm flipped}&\simeq 
    9.7 \times 10^{35} \times  |(U_l)_{21}|^{-2}
    \biggl(\frac{M_X}{10^{16}~{\rm GeV}}\biggr)^4
     \biggl(\frac{0.0378}{\alpha_5}\biggr)^2 ~{\rm yrs}~.
     \label{tp0mf}
\end{align}
As seen in the above expressions, the proton decay rates depend on the unitary matrix $U_l$ associated with the embedding of the left-handed lepton fields into the $\bar{\bf 5}$ fields (see Eq.~\eqref{assign}). As discussed in Ref.~\cite{Ellis:2020qad}, for a light neutrino mass matrix that has a hierarchical structure that is either normally ordered (NO) or inversely ordered (IO), the relevant matrix elements of $U_l$ may be approximated by 
\begin{align}
 (U_l)_{11}
&\simeq 
\begin{cases}
 (U_{\rm PMNS}^*)_{11} = c_{12} c_{13} &\qquad \text{(NO)} \\[5pt]
 (U_{\rm PMNS}^*)_{13} = s_{13} e^{i\delta - i\frac{\alpha_3}{2}} &\qquad \text{(IO)}
\end{cases}
~, \label{eq:ul11}\\[3pt]
 (U_l)_{21}
&\simeq
\begin{cases}
 (U_{\rm PMNS}^*)_{21} =  -s_{12} c_{23} -c_{12} s_{23} s_{13}
 e^{-i\delta} &\qquad \text{(NO)} \\[5pt]
 (U_{\rm PMNS}^*)_{23} = s_{23} c_{12} e^{-i\frac{\alpha_3}{2}} &\qquad \text{(IO)}
\end{cases}
~, \label{eq:ul21}
\end{align}
where $c_{ij} \equiv \cos \theta_{ij}$ and $s_{ij} \equiv \sin \theta_{ij}$ are the mixing angles, $\delta$ is the Dirac phase, and $\alpha_3$ is a Majorana phase in the PMNS matrix. We use these relations in the following calculation, in which case the ratio of the $\mu^+ \pi^0$ and $e^+ \pi^0$ partial decay widths is predicted to be 
\begin{equation}
     \frac{\Gamma (p\to  \mu^+ \pi^0 )_{\rm
 flipped}}{\Gamma (p\to  e^+ \pi^0)_{\rm flipped}}
\simeq 
\begin{cases}
 0.10 & \mathrm{(NO)} \\ 
 22.9 & \mathrm{(IO)} 
 \label{pdecayIO}
\end{cases}
~.
\end{equation}
Both of these values are much larger than the prediction in  conventional supersymmetric SU(5), which is $\simeq 0.008$. The rate for
$p \to \mu^+ \pi^0$ in the IO scenario is expected to be similar to that for $p\to e^+ \pi^0$ in the NO scenario, and the sensitivity
of Hyper-Kamiokande to the $\mu^+ \pi^0$ final state is expected to be
similar to that to the $e^+ \pi^0$ final state~\cite{HK}.\\

\subsection{Error Estimates}
\label{sec:errors}

We provide in this Section a brief derivation of the estimates of dominant errors in the proton lifetime. 
We look at two contributions to these error estimates, namely the effect of the uncertainty in $g_3$ on the mass of $M_X$ 
and the effects of the uncertainties in the matrix elements.  To determine the effect of $g_3$, we look at the dependence of 
\begin{eqnarray}
M_X=g_5V=\frac{g_5^{1/2} M_{GUT} }{2(\lambda_4\lambda_5)^{1/4}}\exp\left[2\pi^2\left(\frac{1}{g_2^2(M_{GUT})}-\frac{1}{g_3^2(M_{GUT})}\right)\right]
\label{MX}
\end{eqnarray}
on $g_3$, ignoring the $g_3$ dependence of $g_5$. We have checked numerically that this can safely be ignored. 
Since (\ref{MX}) is determined at the scale $M_{GUT}$, the scale at which $g_3$ and $g_2$ unify, the variation of the exponential due to the error in $g_3$
has no effect. This means that the leading-order dependence of $M_X$ on $g_3$ is due to
the change in the matching scale $M_{GUT}$. To approximate this effect on the lifetime, 
we need the one-loop expressions for the gauge couplings $g_2,g_3$:
\begin{eqnarray}
&&\frac{1}{g_2^2(M_{GUT})}=\frac{1}{g_2^2(M_Z)} -\frac{1}{8\pi^2}\ln\left(\frac{M_{GUT}}{M_{Z}}\right)\, , \\
&& \frac{1}{g_3^2(M_{GUT})}=\frac{1}{g_3^2(M_Z)} +\frac{3}{8\pi^2}\ln\left(\frac{M_{GUT}}{M_{Z}}\right) ~,
\end{eqnarray}
where $M_{GUT}$ is a function of $g_3$ defined by the relation $g_2(M_{GUT})=g_3(M_{GUT})$. The $g_3$ dependence of $M_{GUT}$ is given by the following expression 
\begin{eqnarray}
M_{GUT}=M_Z \exp\left[2\pi^2\left(\frac{1}{g_2^2(M_Z)}-\frac{1}{g_3^2(M_Z)}\right)\right]~.
\end{eqnarray}
The estimated error in $M_X$ is then
\begin{eqnarray}
\Delta M_X= M_X \frac{\pi}{2}\frac{\Delta \alpha_s}{\alpha_s^2} \, ,
\end{eqnarray}
where $\alpha_s$ is the strong coupling constant, and $\Delta \alpha_s$ is its uncertainty. 
Since the proton lifetime scales as $M_X^4$, we have~\footnote{This uncertainty is larger than that in the conventional SU(5) by a factor $9/2$ (see Ref.~\cite{Ellis:2019fwf}). } 
\begin{eqnarray}
\Delta_{g_3}\tau_{p\to \pi (e,\mu)}=\tau_{p\to (e,\mu)} 2\pi \frac{\Delta \alpha_s}{\alpha_s^2} \, .
\end{eqnarray}

Estimating the error in the lifetime due to the uncertainties in  the  matrix elements is straightforward,
as the lifetime scales as the inverse of the matrix element squared. This leads to an error estimate of 
\begin{eqnarray}
\Delta_M \tau_{p\to \pi (e,\mu)} =2\tau_{p\to \pi (e,\mu)} \frac{\Delta M_i}{M_i} \, ,
\end{eqnarray}
where $M_i$ denotes the matrix elements and $\Delta M_i$ is their uncertainties. 

The total error estimate is then
\begin{eqnarray}
\Delta \tau_{p\to \pi (e,\mu)} =\sqrt{ \Delta_{g_3}\tau_{p\to \pi (e,\mu)}^2+ \Delta_M \tau_{p\to \pi (e,\mu)}^2} \, .
\end{eqnarray}

\section{Results}
\label{sec:results}

\subsection{Universal Boundary Conditions}
\label{sec:universal}

\begin{figure}[!ht]
\centering
\includegraphics[width=8.5cm]{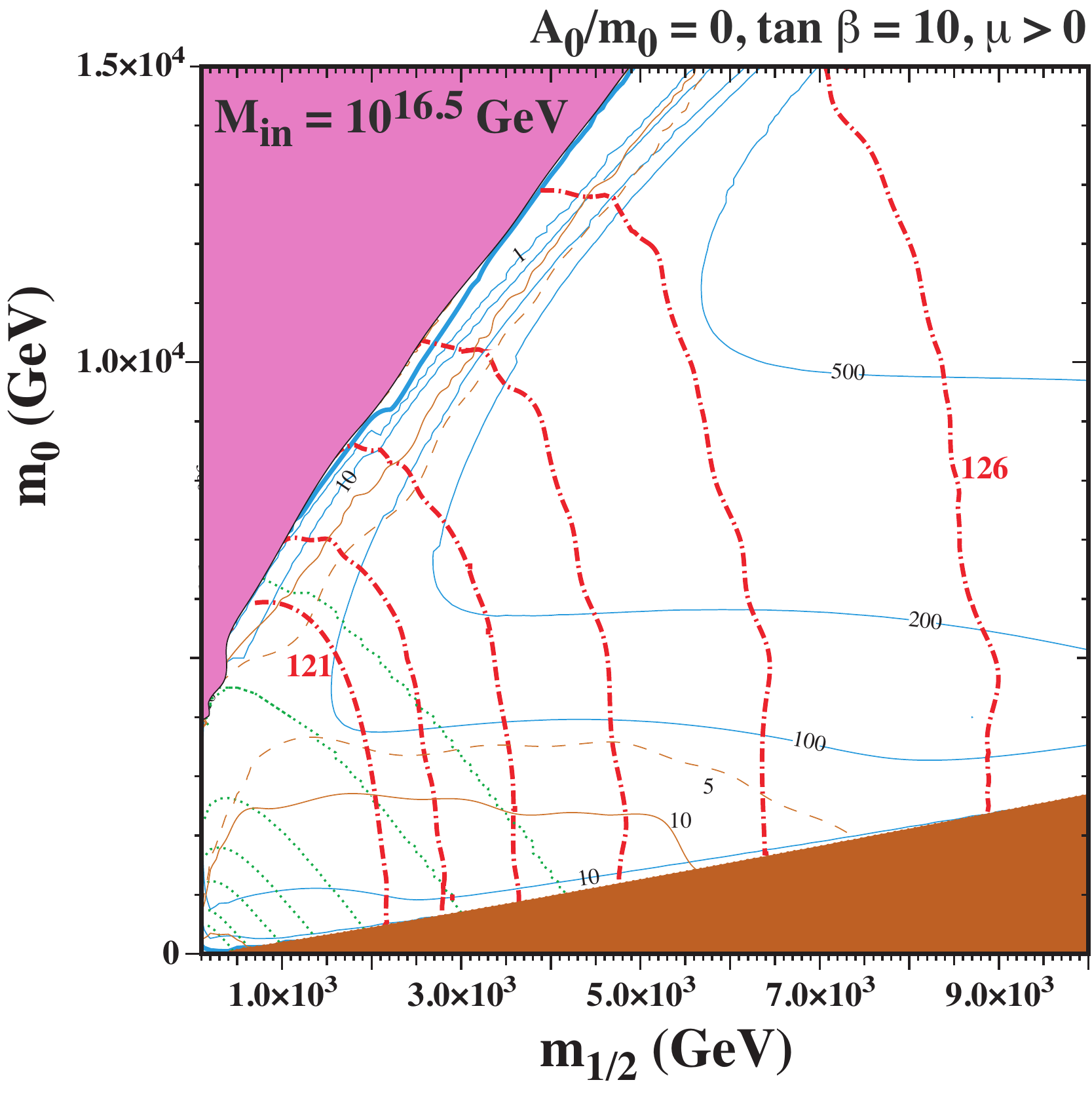}
\caption{\it A representative $(m_{1/2}, m_0)$ plane in the flipped SU(5) GUT model, with $M_{in} = 10^{16.5}$~GeV,
$\tan \beta = 10, A_0 = 0, \lambda_4 = 0.3, \lambda_5 = 0.1$ and $\lambda_6 = 0.0001$. Regions with a stau LSP
and without electroweak symmetry breaking are shaded brown and pink, respectively. 
The red dot-dashed curves are contours of the Higgs mass from 121 - 126 GeV in intervals of 1 GeV, as calculated using {\tt FeynHiggs~2.18.1}.  The solid blue lines are contours of the relic density $\Omega_{LSP} h^2 = 0.1, 1, 10, 100, 200, 500$.  A thicker contour with $\Omega_{LSP} h^2 = 0.1$ is visible near the border of the no-electroweak symmetry breaking region, i.e., the focus-point region.  The solid brown lines are where the central value of the $p \to e^+ \pi^0$ lifetime is $10^{36}$~years (labelled 10 in units of $10^{35}$ years) and the dashed brown lines are where $\tau_p - \sigma_{\tau_p} = 5\times 10^{35}$ years.
The green dotted lines are contours of $\Delta a_\mu = 1, 2, 5,10, 20, 50, 100 \times 10^{-11}$, increasing as $m_{1/2}$ and $m_0$ decrease.}
\label{fig:plane}
\end{figure}

We examine first a selection of $(m_{1/2}, m_0)$ planes when universal boundary conditions are applied
at a high input scale $M_{in} > M_{GUT}$. Our baseline plane shown in
Fig.~\ref{fig:plane} is similar to 
that considered in~\cite{FSU5Cosmo} 
with $\tan \beta = 10$, $A_0 = 0$, $M_{in} = 10^{16.5}$ GeV,
{\boldmath $\lambda$} $\equiv (\lambda_4, \lambda_5) 
= (0.3, 0.1)$, $\lambda_6 = 10^{-4}$, and 
$\mu > 0$.~\footnote{This and subsequent planes are generally not 
sensitive to $\lambda_6$. This coupling enters into the neutrino mass matrix, and our choice here corresponds roughly to the example in \cite{FSU5Cosmo}.} The pink shaded region at 
large $m_0 \gg m_{1/2}$ is excluded by the absence 
of a consistent electroweak vacuum, and the brown shaded region 
where $m_{1/2} \gg m_0$ is excluded because 
the lighter stau is the LSP and/or tachyonic. The red 
dot-dashed lines are contours of constant Higgs masses 
between $m_h = 121$ and 126 GeV in intervals of 1 GeV, as calculated using {\tt FeynHiggs 2.18.10}~\cite{FH}. {We
consider calculated values of $m_h \in (122, 128)$~GeV to
be consistent with the measured value within 
conservative calculational uncertainties.}

The solid blue contours in Fig.~\ref{fig:plane}
show values of the LSP relic density, $\Omega_\chi h^2$, as labeled, as calculated assuming that the
Universe expands adiabatically. 
The contour for $\Omega_\chi h^2 =0.1$,
corresponding to the measured dark matter density,
appears as a thick blue curve near the pale blue shaded area, 
and corresponds to the focus-point region~\cite{fp}.
There is also a short contour with $\Omega_\chi h^2 =0.1$ just 
above the stau-LSP region with $m_{1/2} \lesssim 1$ TeV~\cite{efo}
that is almost invisible. As mentioned previously, the generation of
a large amount ($\mathcal{O}(10^4$) ) of entropy in the early Universe due to the late decay of the flat direction responsible for the breaking of FSU(5) is a generic
feature of FSU(5) cosmology~\cite{egnno2,egnno3,egnno4,FSU5Cosmo}, so we do not interpret this
adiabatic calculation of $\Omega_\chi h^2$ as a necessary
constraint. Indeed when accounting for the late entropy production, we expect that parameters yielding $\Omega h^2 \sim 100-1000$ would correspond better to the present relic density $\Omega h^2 \simeq 0.1$.  (See \cite{Giblin:2017wlo,Allahverdi:2021grt} for related work.)

In addition, we show in Fig.~\ref{fig:plane} 
as the solid brown curve the contour where
$\tau_p (p\to e^+ \pi^0) = 10^{36}$ yrs, as calculated
assuming normal ordering (NO) of the neutrino masses. 
This line appears at $m_0 \approx 2.5$ TeV and also runs roughly 
parallel to the focus-point strip. The proton lifetime varies 
slowly across this plane, in general, and is always within the
range 5 -- 20 $\times
10^{35}$ yrs, beyond the foreseen experimental reach~\cite{HK}.  The brown dashed contour corresponds to 
$\tau_p - \sigma_{\tau_p} = 5 \times 10^{35}$~yrs, illustrating
the effect of the uncertainty in the calculation of $\tau_p$
discussed in the previous Section.~\footnote{The proton lifetime 
for $p \to \mu^+ \pi^0$ assuming NO can be obtained by comparing Eqs.~(\ref{tp0ef}) and (\ref{tp0mf}). The lifetime assuming IO can be found using Eq.~(\ref{pdecayIO}).}  Clearly,
the relatively long proton lifetime in this plane makes detection difficult. 
Finally, we show a series of curves of constant $\Delta a_\mu$ as indicated in the caption,
with the largest values appearing at small $m_{1/2}$ and $m_0$.
We note that all values of $\Delta a_\mu > 2 \times 10^{-11}$
appear for values of $m_h < 122$~GeV, outside the range that we
consider compatible with the measured value of $m_h$. 

\begin{figure}[!ht]
\centering
\includegraphics[width=7cm]{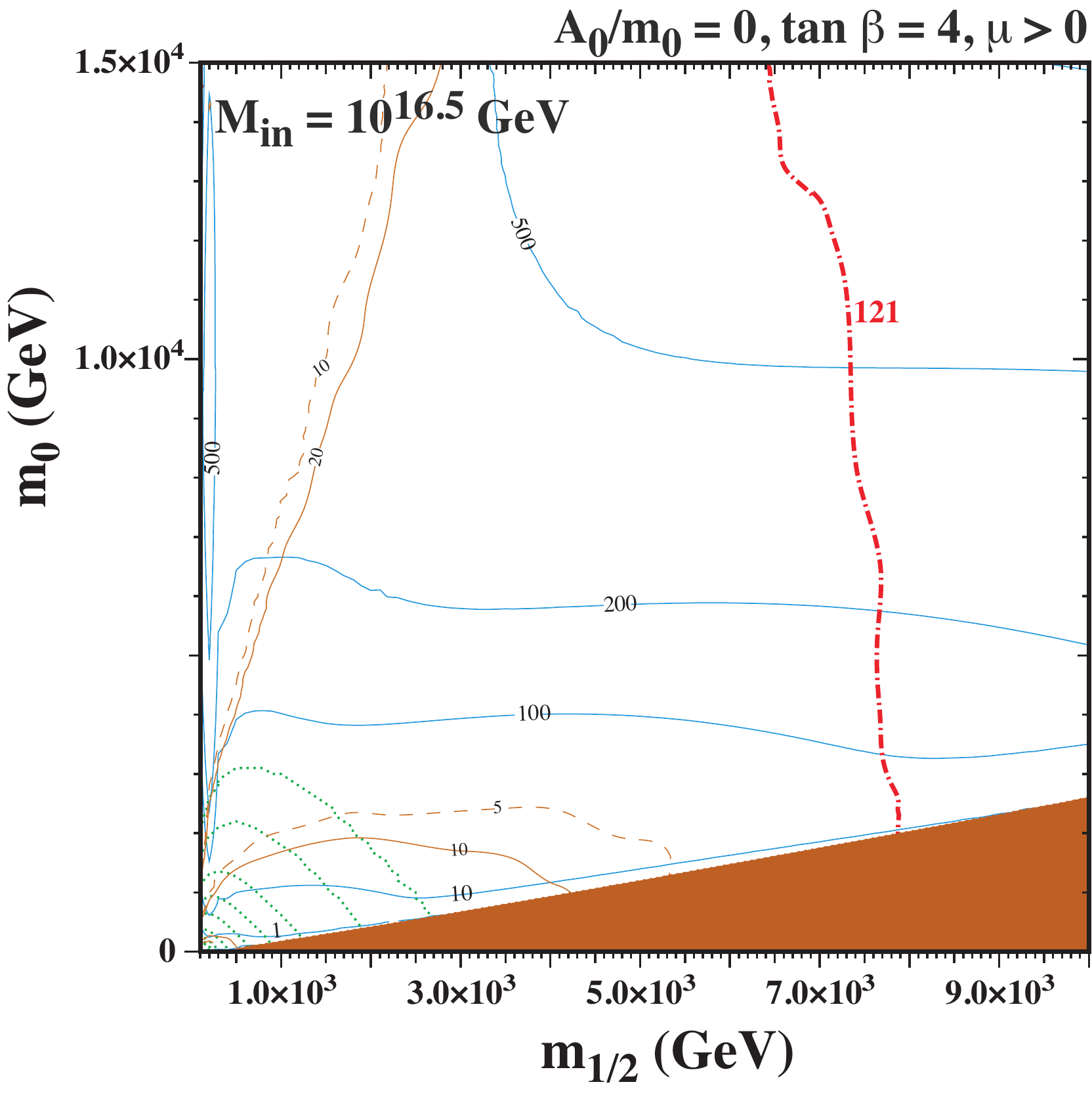}
\hspace{0.25in}
\includegraphics[width=7cm]{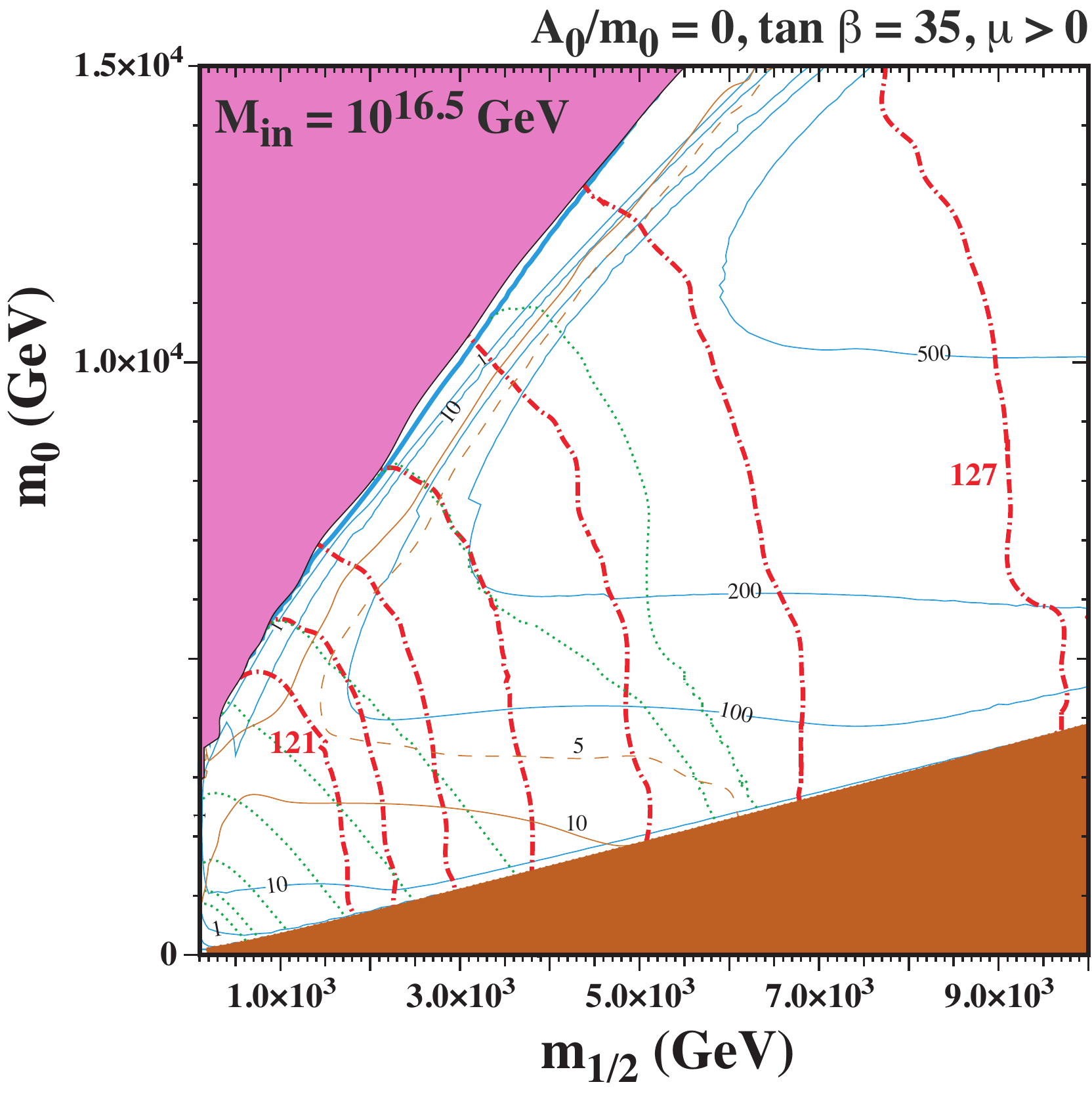}
\caption{\it Representative $(m_{1/2}, m_0)$ planes in the FSU(5) GUT model.
The parameters as the same as in Fig.~\ref{fig:plane} except that
$\tan \beta = 4$ in the left panel and $\tan \beta = 35$ in the right panel. 
Regions with a stau LSP
and without electroweak symmetry breaking are shaded brown and pink,
respectively. The  contours are as in Fig.~\ref{fig:plane}, with the
addition of contours for $\tau_p = 2 \times 10^{36}$ years and $\tau_p - \sigma_{\tau_p} = 10^{36}$ years in the left panel.}
\label{fig:planestb}
\end{figure}

In Fig.~\ref{fig:planestb}, we compare analogous planes with different values of 
$\tan \beta$.  In the left panel, $\tan \beta = 4$, while in the right panel 
$\tan \beta = 35$. All other fixed parameters are the same as in Fig.~\ref{fig:plane}.
For $\tan \beta = 4$, The Higgs mass is always less than 122 GeV, and only the 
$m_h = 121$~GeV contour appears. The region without consistent electroweak symmetry
breaking is pushed out beyond the range of the plot, so there is no visible
focus-point region, and entropy production is required throughout the displayed
plane. Compared to Fig.~\ref{fig:plane}, the proton lifetime is longer and the 
values of $\Delta a_\mu$ are smaller for given values of $(m_{1/2}, m_0)$. 
In contrast, for the larger value of $\tan \beta = 35$ shown in the right panel
of Fig.~\ref{fig:planestb}, the region without electroweak symmetry breaking
extends to lower values of $m_{1/2}$ and $m_0$, and the Higgs mass is higher,
rising beyond 127~GeV in this plane.
Though the proton lifetime is somewhat lower than in Fig.~\ref{fig:planestb}, 
it is still very large. On the other hand, values of $\Delta a_\mu$
are larger than in Fig.~\ref{fig:planestb}, and reach $10^{-10}$ for
$m_h > 122$ GeV. 

\begin{figure}[!ht]
\centering
\includegraphics[width=7cm]{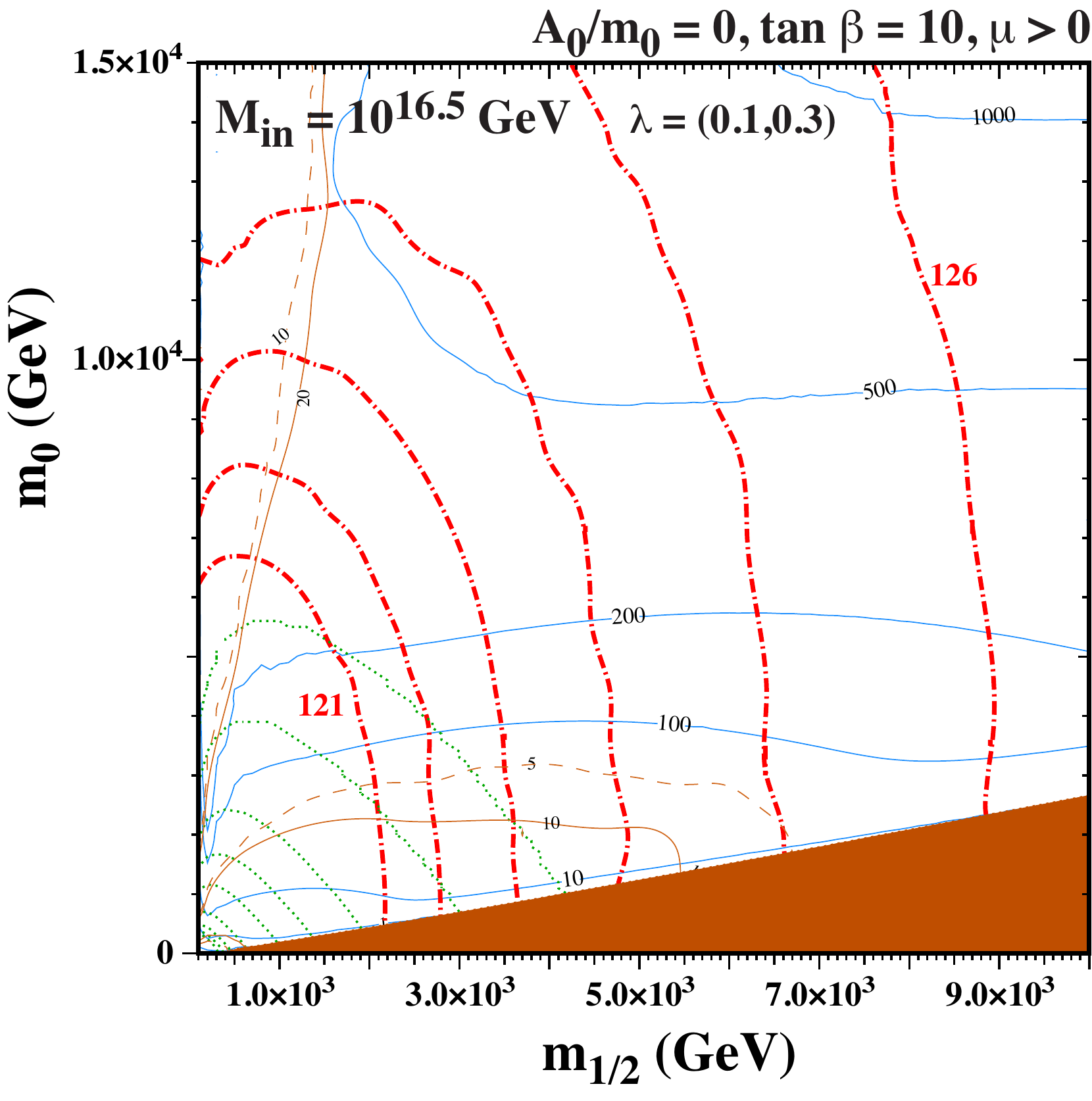}
\hspace{0.25in}
\includegraphics[width=7cm]{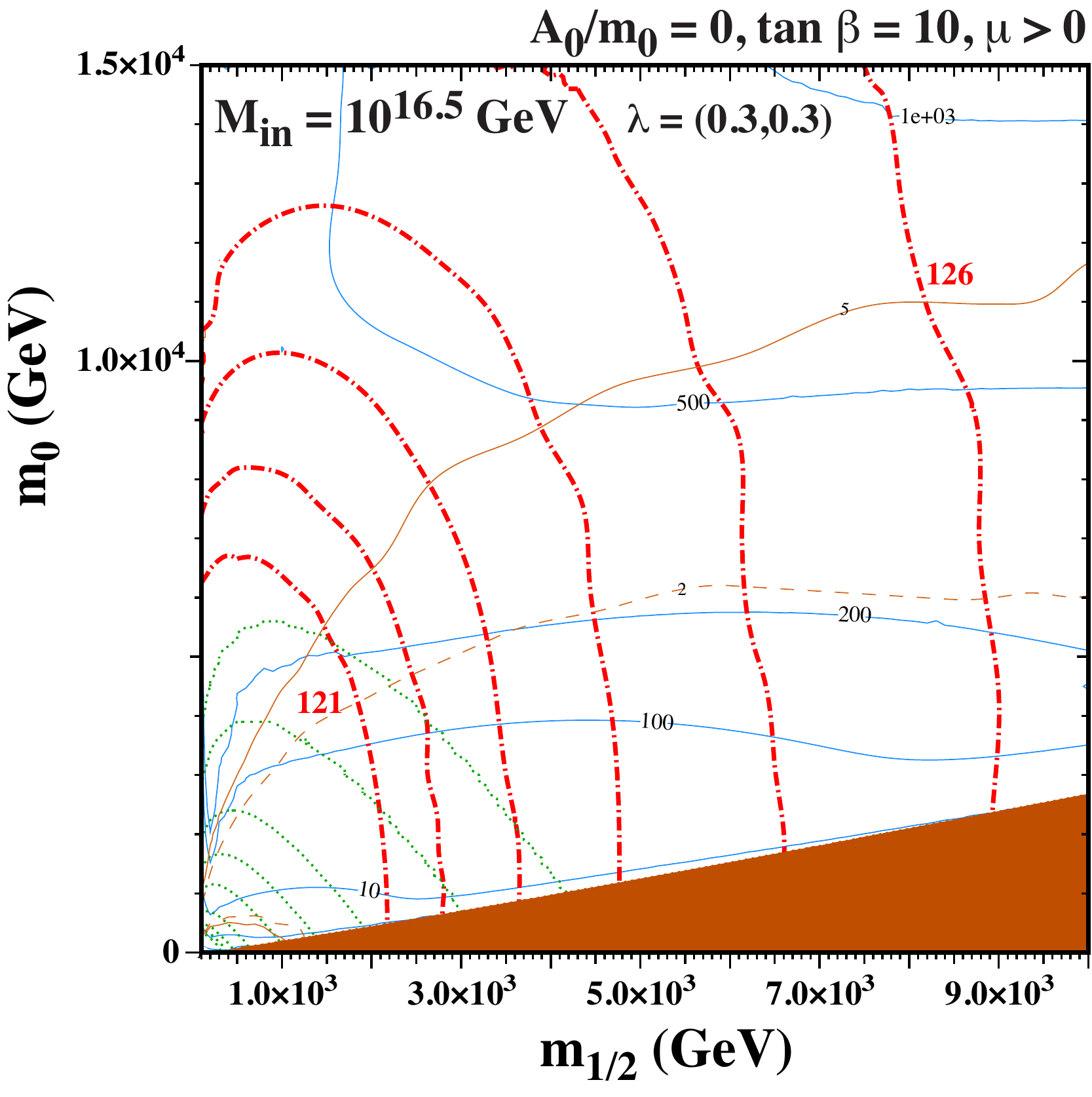}\\
\includegraphics[width=7cm]{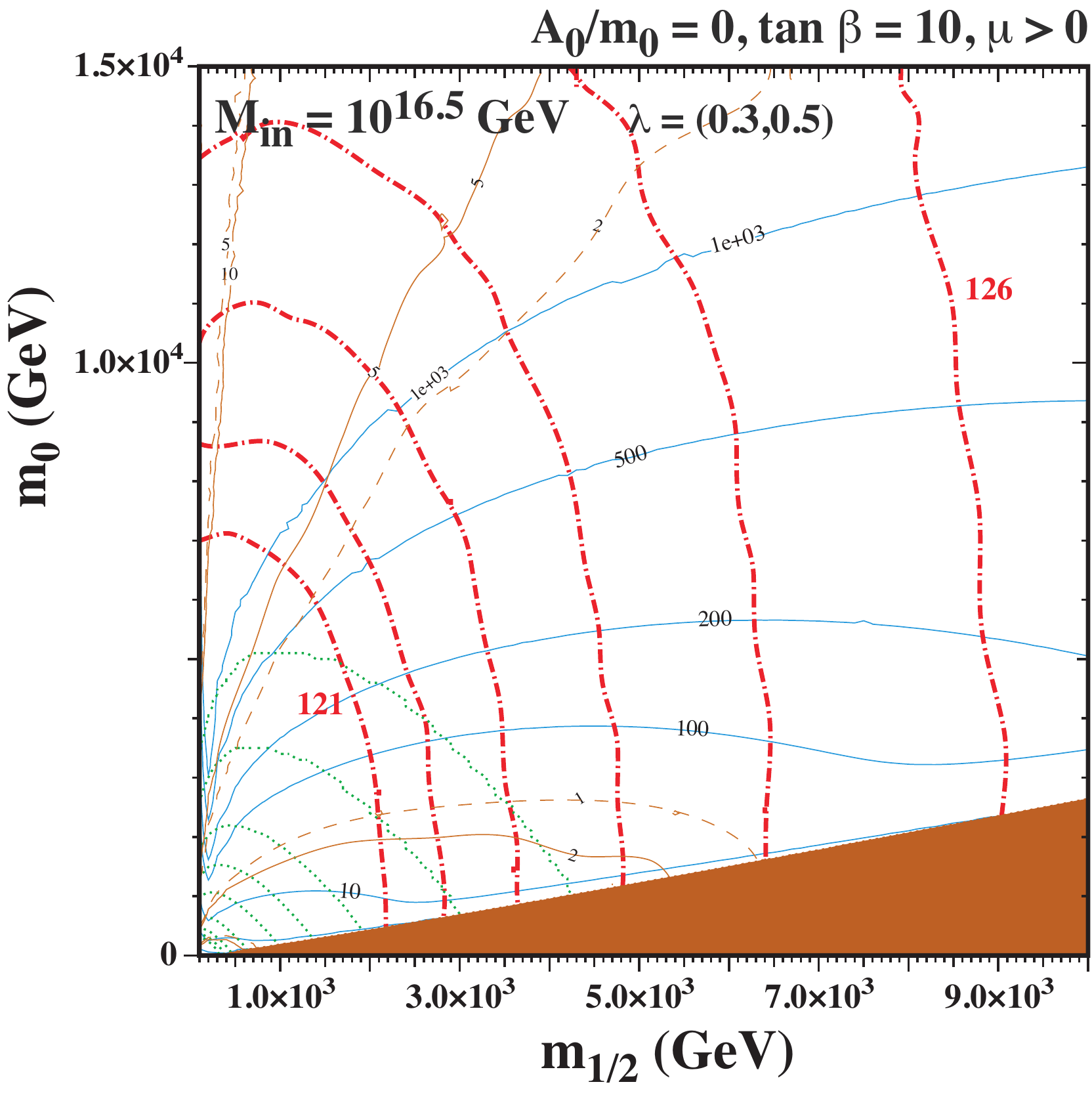}
\hspace{0.25in}
\includegraphics[width=7cm]{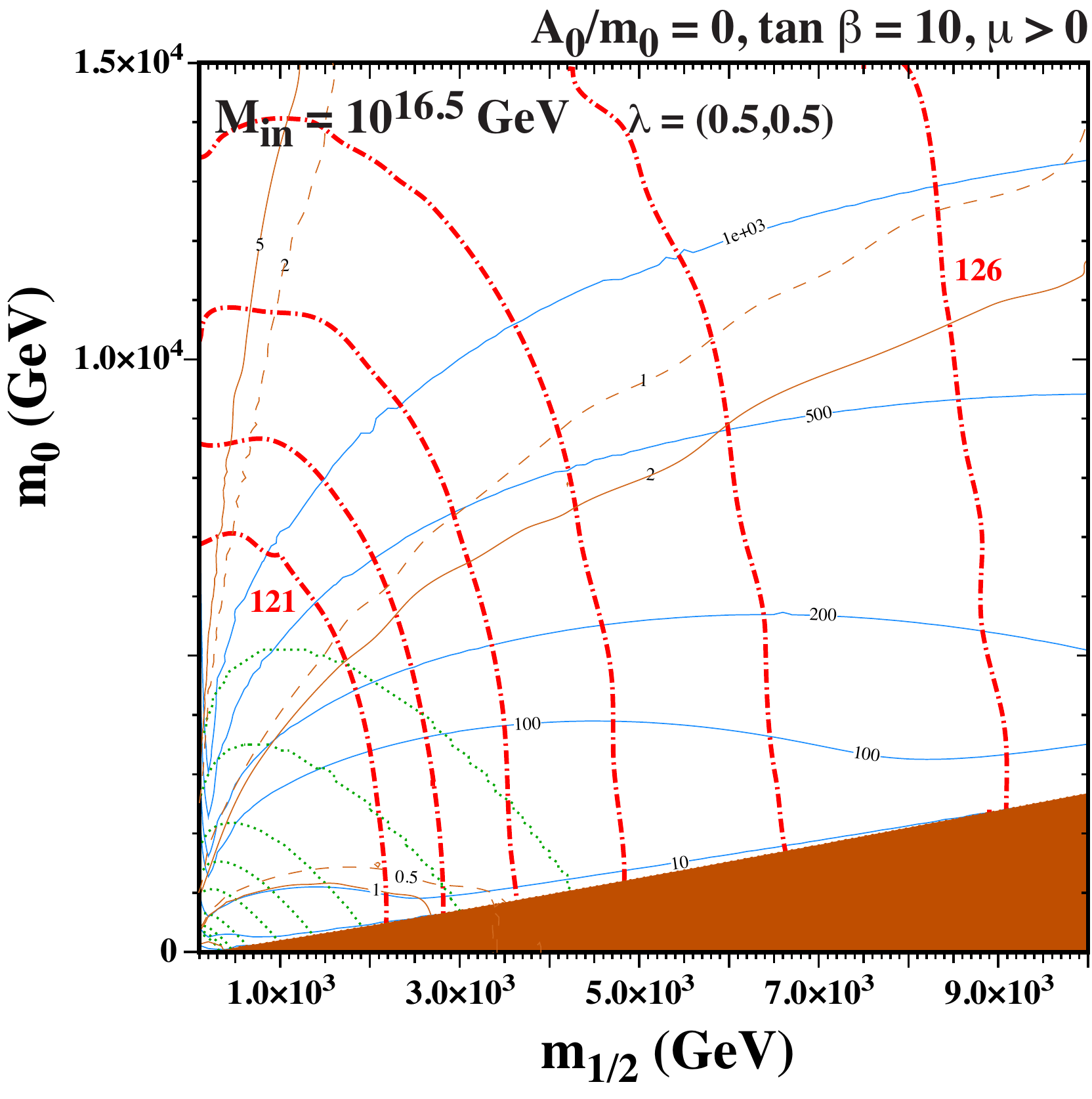}\\
\caption{\it Representative $(m_{1/2}, m_0)$ planes in the FSU(5) GUT model.
The fixed parameters as the same as in Fig.~\ref{fig:plane}, except for
{\boldmath{$\lambda$}} = (0.1,0.3) in the top left panel,
{\boldmath{$\lambda$}} = (0.3,0.3) (top right), 
{\boldmath{$\lambda$}} = (0.3,0.5) (bottom left), and
{\boldmath{$\lambda$}} = (0.5,0.5) (bottom right).
Regions with a stau LSP are shaded brown. The  contours are as in Fig.~\ref{fig:plane}}
\label{fig:planesL}
\end{figure}

We explore the dependence on {\boldmath $\lambda$} in Fig.~\ref{fig:planesL}.
Keeping the other parameters fixed to the values used in
Fig.~\ref{fig:plane}, we take {\boldmath $\lambda$} = (0.1,0.3) in the
upper left panel of Fig.~\ref{fig:planesL}, (0.3,0.3) (upper right), 
(0.3,0.5) (lower left), and (0.5,0.5) (lower right). 
None of the planes displays a constraint from electroweak symmetry breaking. 
This is tied to the larger value of $\lambda_5$ used here 
($\ge 0.3$ vs the value of 0.1 used in Fig.~\ref{fig:plane}). 
The Higgs mass and muon magnetic moment are relatively insensitive to 
{\boldmath $\lambda$}. However, we see from Eq.~(\ref{MX})
that the $X$ gauge boson mass is inversely proportional to 
$\sqrt{\lambda_4 \lambda_5}$, so that increasing this product leads to a 
smaller mass and hence a shorter proton lifetime. Thus, whereas
the lifetime for {\boldmath $\lambda$} = (0.1,0.3) is similar to the 
lifetime with (0.3,0.1), for {\boldmath $\lambda$} = (0.3,0.3)
we see a mean lifetime contour of $5 \times 10^{35}$ years and a 
1$\sigma$-reduced lifetime of $2 \times 10^{35}$ years running through 
the upper right panel. When the product $\sqrt{\lambda_4 \lambda_5}$ is 
further increased, as in the lower left panel with 
{\boldmath $\lambda$} = (0.3,0.5), we see
lifetime contours of 2, 5, and 10 $\times 10^{35}$ years, and 
reduced lifetime contours of 1, 2, and 5 $\times 10^{35}$ years. Finally,
in the lower right panel with {\boldmath $\lambda$} = (0.5,0.5), we find 
proton lifetime contours of 1, 2, and 5 $\times 10^{35}$ years, 
and 1$\sigma$-reduced 
lifetimes of 0.5, 1, and 2 $\times 10^{35}$ years.
We recall that lifetimes $\lesssim 10^{35}$~years
open up the possibility of detection in the upcoming Hyper-Kamiokande experiment~\cite{HK}.
Finally, we note that higher values of $\lambda_{4,5}$ lead to problems in the running of the 
RGEs down from $M_{in}$ to $M_{GUT}$.

\begin{figure}[!ht]
\centering
\includegraphics[width=7cm]{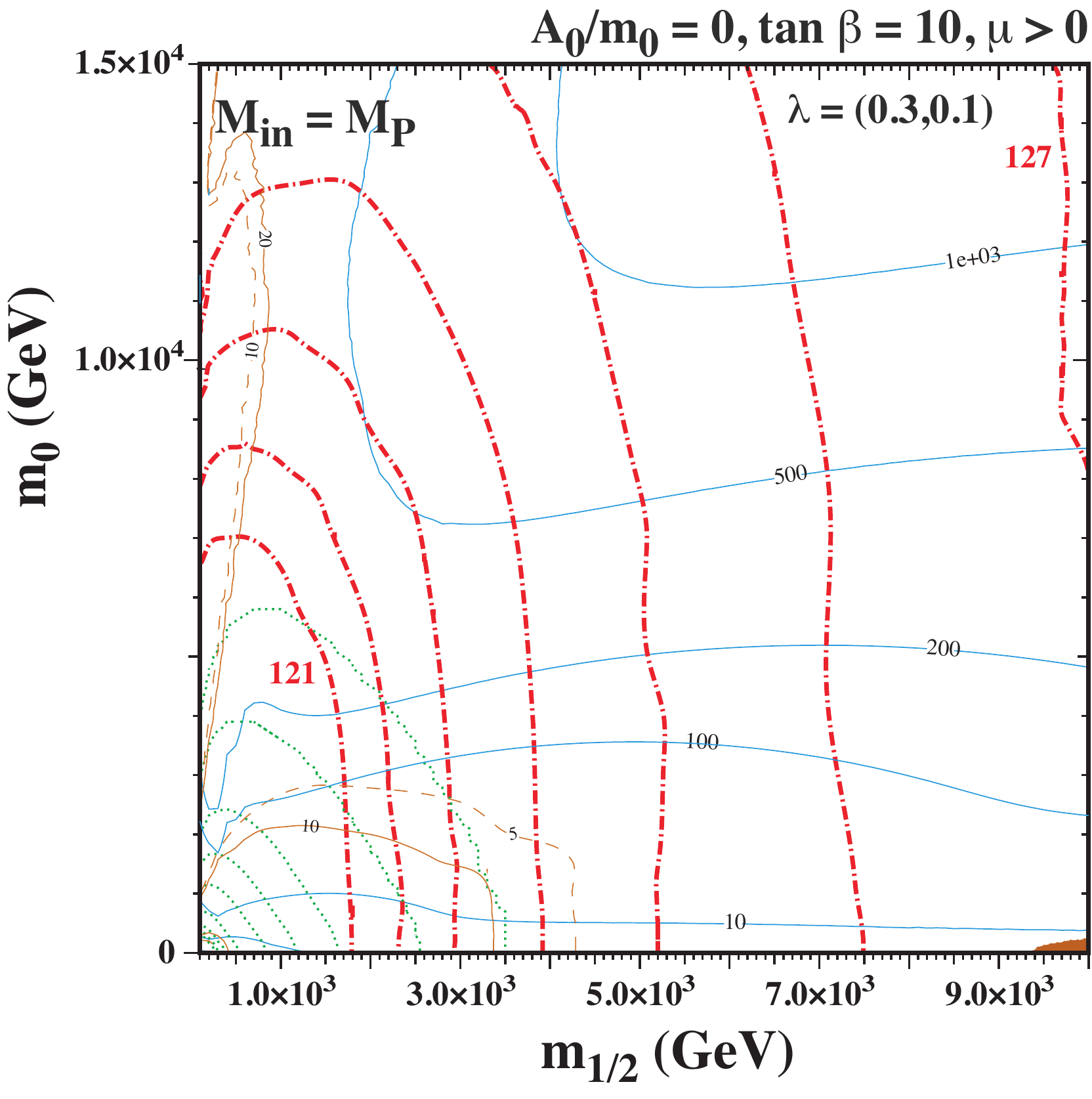}
\hspace{0.25in}
\includegraphics[width=7cm]{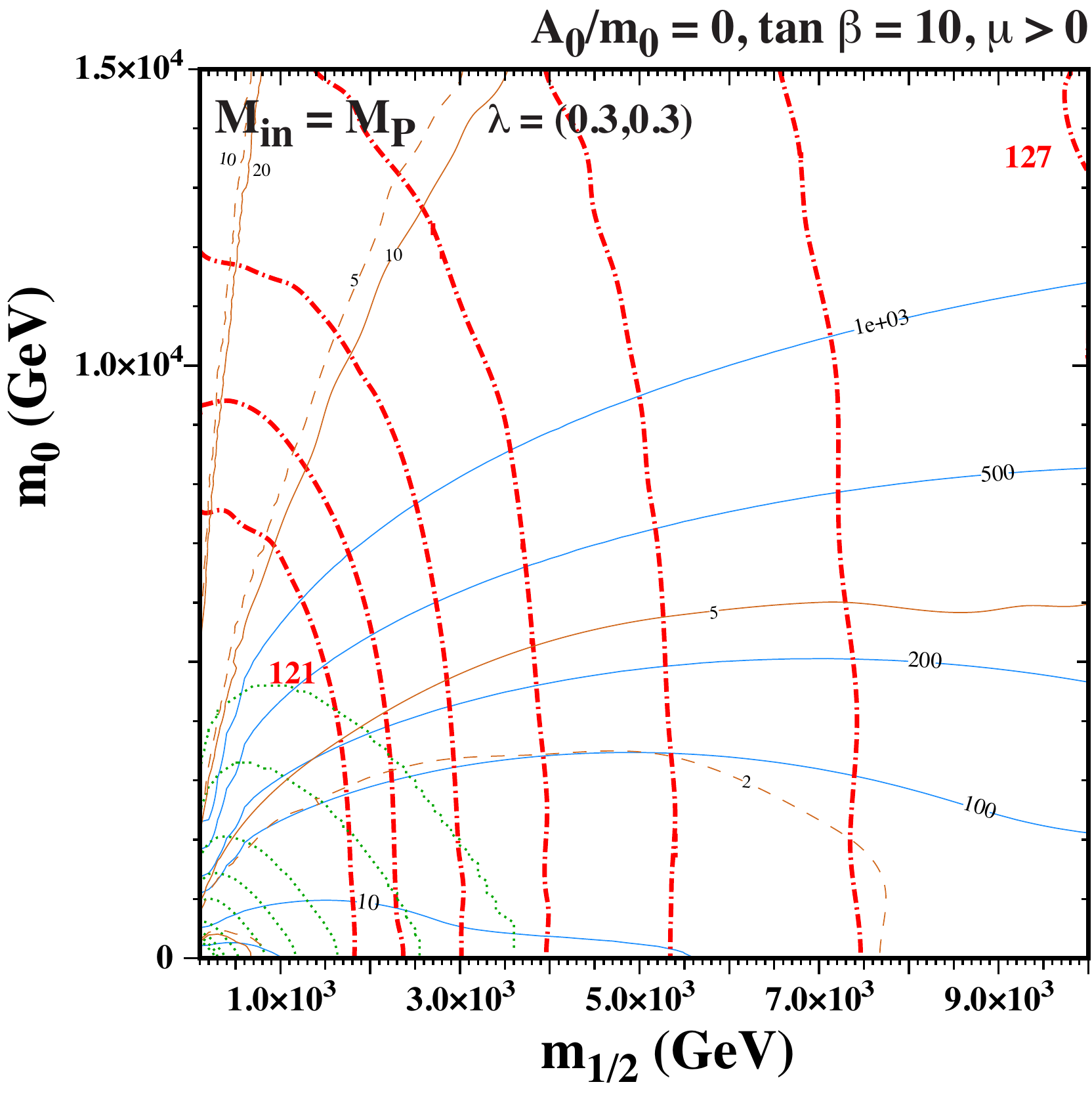}
\caption{\it Representative $(m_{1/2}, m_0)$ planes in the FSU(5) GUT model. 
The fixed parameters as the same as in Fig.~\ref{fig:plane}, except for
$M_{in} = M_P$ and  {\boldmath{$\lambda$}} = (0.3,0.1) in the left
panel and {\boldmath{$\lambda$}} = (0.3,0.3) (right).
The  contours are as in Fig.~\ref{fig:plane}.}
\label{fig:planeMin}
\end{figure}

The dependence on $M_{in}$ is considered in Fig.~\ref{fig:planeMin}. The left panel
assumes the same parameter values as in Fig.~\ref{fig:plane}, with the exception of
$M_{in}$, which is now set at the Planck scale. Electroweak symmetry breaking 
occurs throughout the plot, and the stau LSP region is pushed to the lower right
corner of the panel. The Higgs mass, proton lifetime and relic density are all slightly larger than in Fig.~\ref{fig:plane}.  In the right panel,
we again take $M_{in} = M_P$ but now with {\boldmath{$\lambda$}} = (0.3,0.3),
which is near its upper limit for this value of $M_{in}$. 

\begin{figure}[!ht]
\centering
\includegraphics[width=7cm]{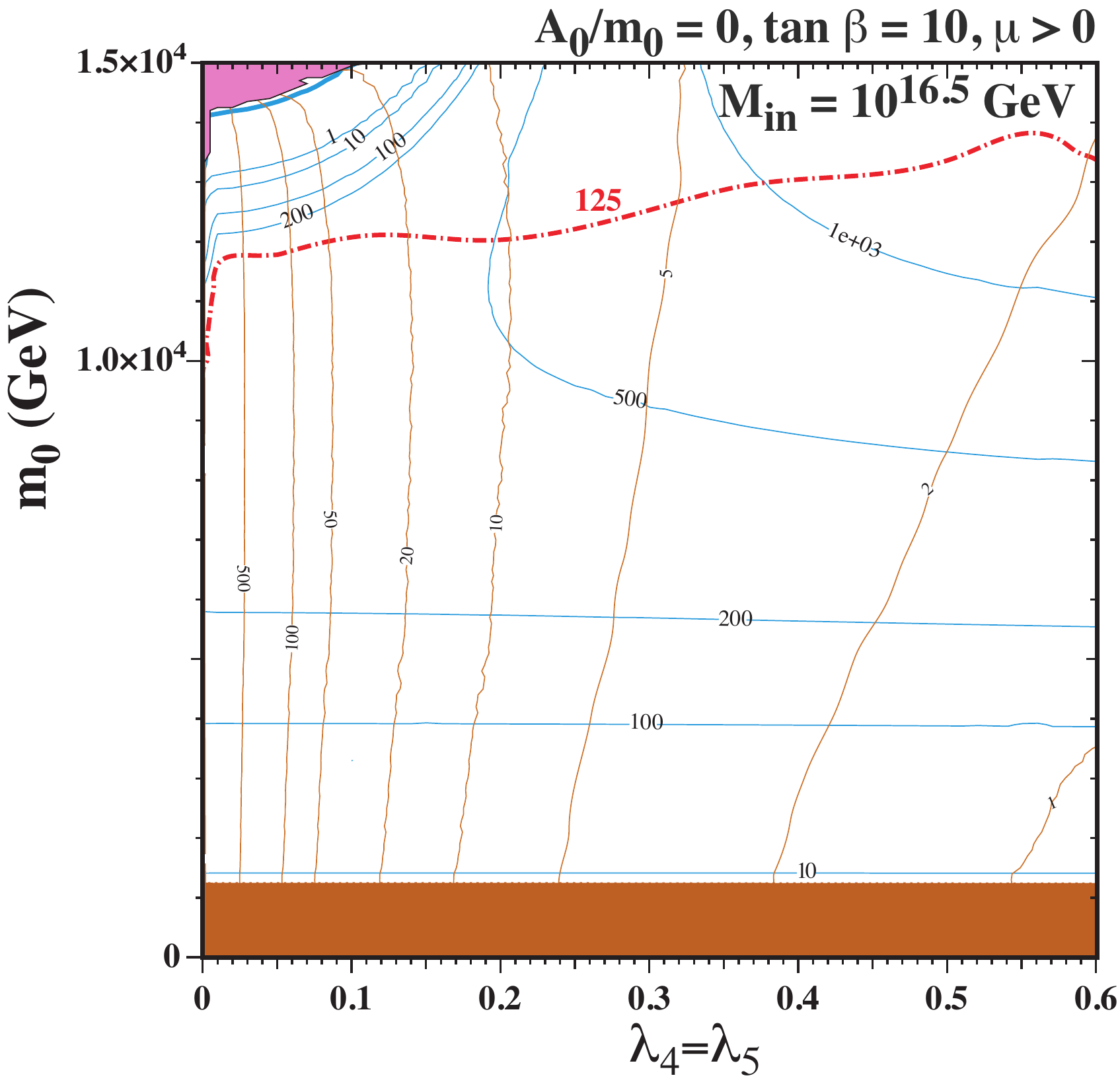}
\hspace{0.25in}
\includegraphics[width=7cm]{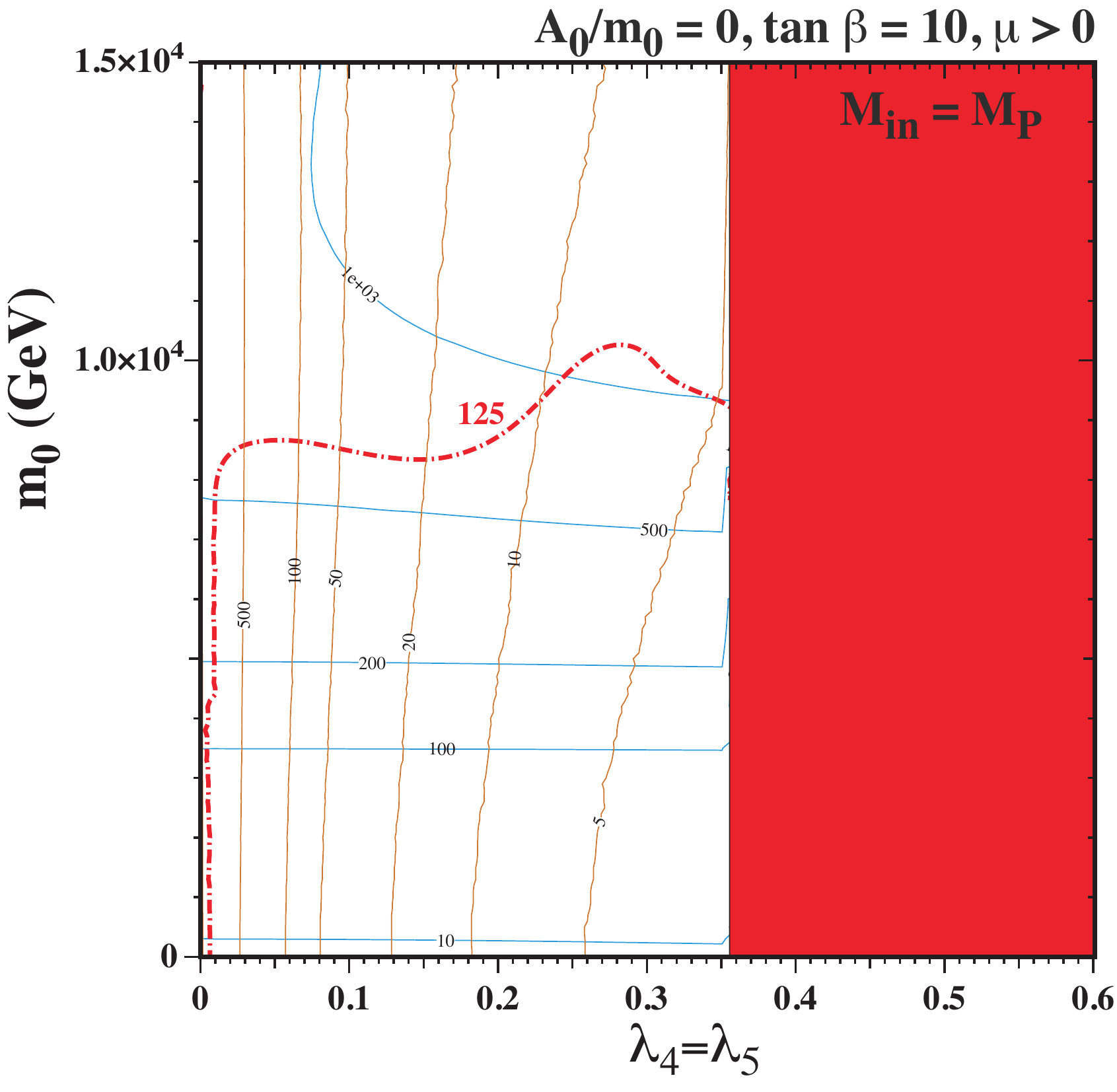}
\caption{\it Representative $(\lambda_4=\lambda_5, m_0)$ planes in the
FSU(5) GUT model with $M_{in} = 10^{16.5}$~GeV ($M_P$) for the left (right)
panel. The brown shaded region in the left panel is excluded because the 
LSP is charged, and there is no electroweak symmetry breaking in the pink
shaded region, while the RGE equations break down in the red shaded band 
at large $\lambda_4=\lambda_5$ in the right panel.
The  contours are as in Fig.~\ref{fig:plane}.}
\label{fig:lamvsm0}
\end{figure}

The relation between the proton lifetime and $\sqrt{\lambda_4 \lambda_5}$ is seen more clearly in Fig.~\ref{fig:lamvsm0},
which shows  a pair of $(\lambda_4=\lambda_5, m_0)$ planes for $m_{1/2} = 5$~TeV, $A_0/m_0 = 0, \tan \beta = 10$, $\mu > 0$, $\lambda_6 = 0.0001$, with
$M_{in} = 10^{16.5}$~GeV  (left panel) and $M_{in} = M_P$ (right panel).
When $M_{in} = 10^{16.5}$~GeV there is a small region in the upper left corner where electroweak symmetry breaking breaks down, which is shaded pink. Bordering this region, the focus-point strip with $\Omega h^2 = 0.1$ is the thick blue contour. Other blue contours correspond to larger values for the relic density, but we re-emphasize
that larger values of $\Omega_\chi h^2$ would be allowed in the context of FSU(5) cosmology, in which substantial
entropy is likely to have been generated in the early Universe. There is a stau LSP in the brown shaded region at low $m_0$ in the left panel.  For $M_{in} = M_P$,
the RGEs break down when $\lambda_4 = \lambda_5 \gtrsim 0.35$, as indicated by the red shading in the right panel.
The red lines are contours of $m_h = 125$ GeV. 
We note that $m_h$ varies slowly across this plane, so this is the only integer mass contour displayed.  Finally, the
solid brown lines are contours of $\tau (p \to e^+ \pi^0)$ in units of $10^{35}$~yrs. We see that values of the proton 
lifetime that are $\lesssim 3 \times 10^{35}$~yrs, and hence potentially accessible to the next generation of experiment,
are found in the right portions of the planes where $\lambda_4=\lambda_5 \gtrsim 0.3$.

\begin{figure}[!ht]
\centering
\includegraphics[width=7cm]{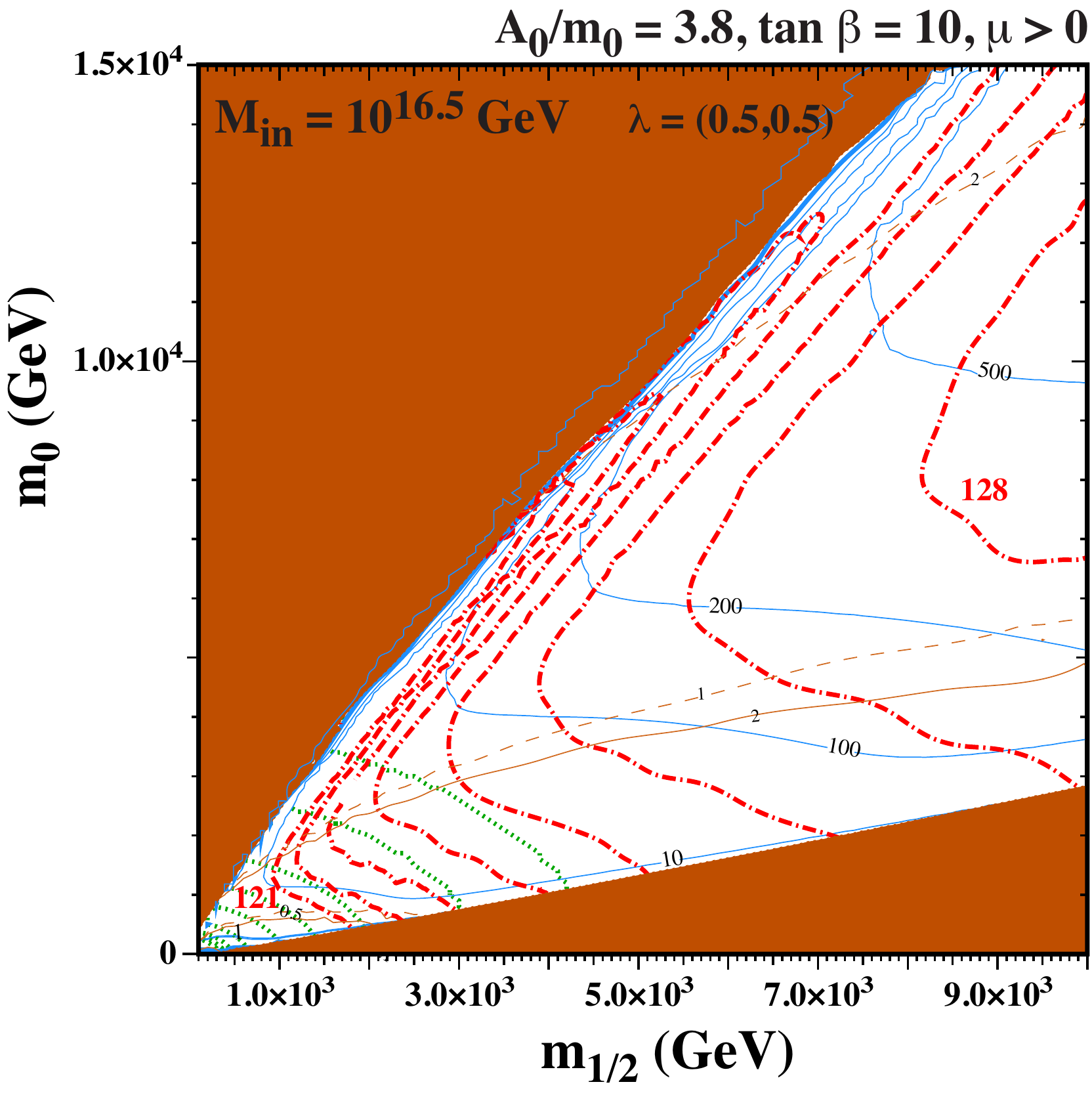}
\hspace{0.25in}
\includegraphics[width=7cm]{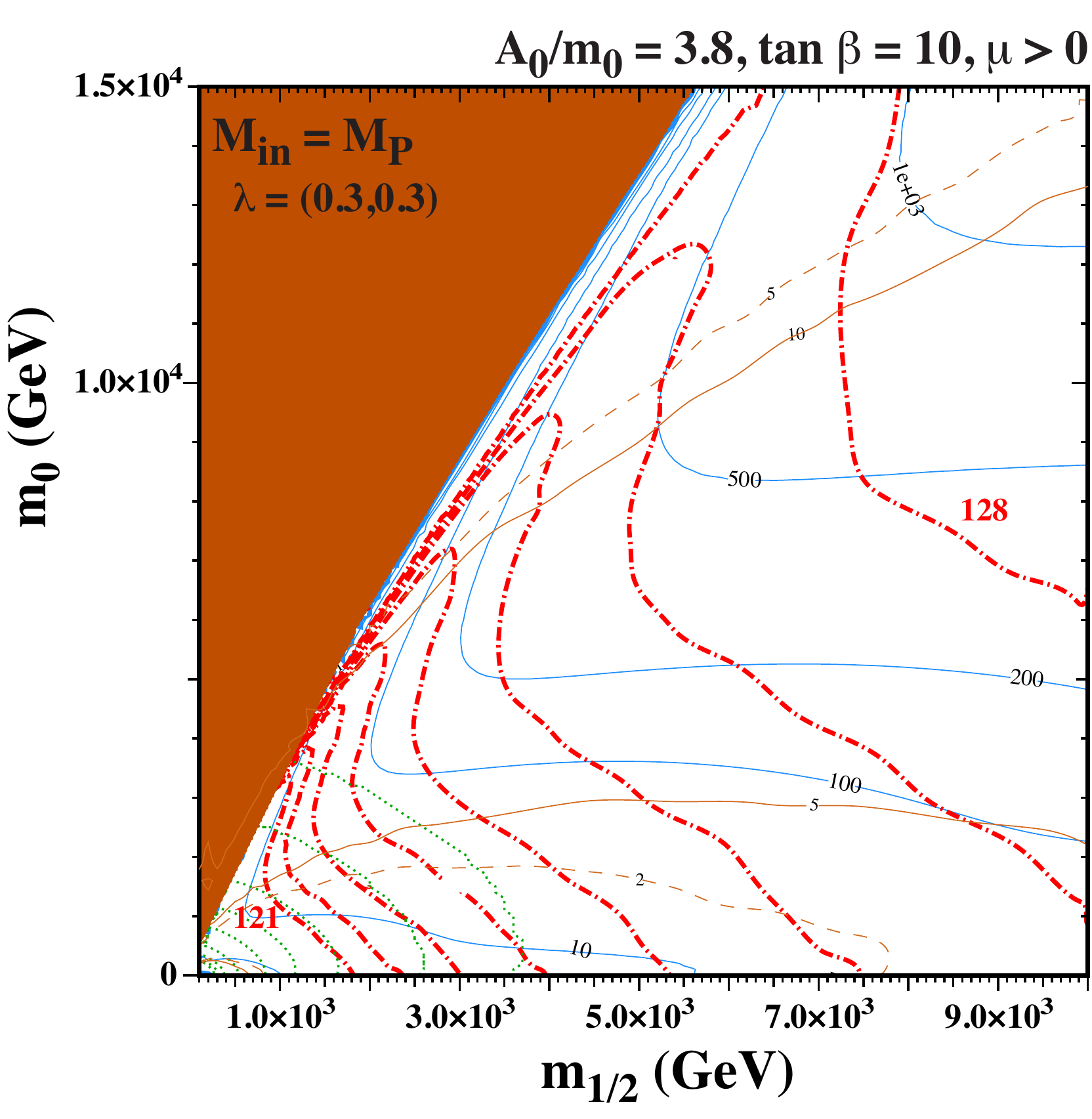}
\caption{\it Representative $(m_{1/2}, m_0)$ planes in the FSU(5) GUT model. 
Parameters as in Fig.~\ref{fig:plane}, except that $A_0/m_0 = 3.8$,
$\lambda_4 = \lambda_5 = 0.5$  (left), and $M_{in} = M_P$
with $\lambda_4 = \lambda_5 = 0.3$,  (right).
Regions with a stau LSP, stop LSP or tachyonic stop/stau are shaded brown.
The  contours are as in Fig.~\ref{fig:plane}.}
\label{fig:stopplanes}
\end{figure}

The parameter planes displayed above have all assumed $A_0 = 0$, and we
present in Fig.~\ref{fig:stopplanes} a pair of planes with non-zero $A_0$, specifically
$A_0 = 3.8 ~m_0$. In the left panel, we take $M_{in} = 10^{16.5}$ GeV and 
{\boldmath $\lambda$} = (0.5,0.5) to minimize the proton lifetime. 
In the right panel, $M_{in} = M_P$ and {\boldmath $\lambda$} = (0.3,0.3). 
These planes exhibit the possible importance of a compressed stop
spectrum, which introduces the possibility of stop coannihilation~\cite{stopco},
and can be compared with Fig.~12c of~\cite{FSU5Cosmo}. The brown shaded
regions where $m_0 > m_{1/2}$ are disallowed because the stop is either the LSP or tachyonic,
and that in the left panel where $m_{1/2} > m_0$ has a stau LSP. As we have seen previously, 
the stau LSP region recedes to larger values of $m_{1/2}$ as $M_{in}$ increases,
and is not visible in the right panel where $M_{in} = M_P$.  We see very large values
of $\Omega_\chi h^2$ in the bulk of the uncoloured region,~\footnote{Which are allowed
in the FSU(5) cosmological scenario described in~\cite{FSU5Cosmo}.} but
there are strips close to the boundaries of the shaded regions
where $\Omega_\chi h^2$ is reduced. Once again, the thick blue shaded contour running along
the stop LSP region corresponds to $\Omega h^2 = 0.1$.

It is important to note that we can find  
$\Omega h^2 = 0.1$ and $m_h = 125$~GeV simultaneously in both panels in
Fig.~\ref{fig:stopplanes}, but at very different values of $(m_{1/2},m_0)$. 
For $M_{in} = 10^{16.5}$~GeV, simultaneity occurs around $(m_{1/2},m_0) \simeq (7,13)$~TeV
where the proton lifetime $\tau_p \approx 5 \pm 3 \times 10^{35}$ years. 
However, for $M_{in} = M_P$, these conditions are both satisfied when 
$(m_{1/2},m_0) \simeq (1.1,3.8)$~TeV.~\footnote{In this case the stop mass is
relatively light ($\simeq 500$~GeV) and nearly degenerate with the bino, and further
detailed studies would be needed to assess it compatibility with LHC constraints.} 
Despite the lower sparticle masses, the
proton lifetime is actually longer here (around $10^{36}$ years),
mainly due to the lower values of $\lambda_{4,5}$ needed to 
ensure 
non-divergent running between $M_P$ and $M_{GUT}$.
In both cases, the contribution to $\Delta a_\mu$ is small ($< 10^{-11}$). 
We stress again, however, that as late-time entropy production is expected 
in this FSU(5) model, most of the displayed plane is viable cosmologically.

\begin{figure}[!ht]
\centering
\includegraphics[width=7cm]{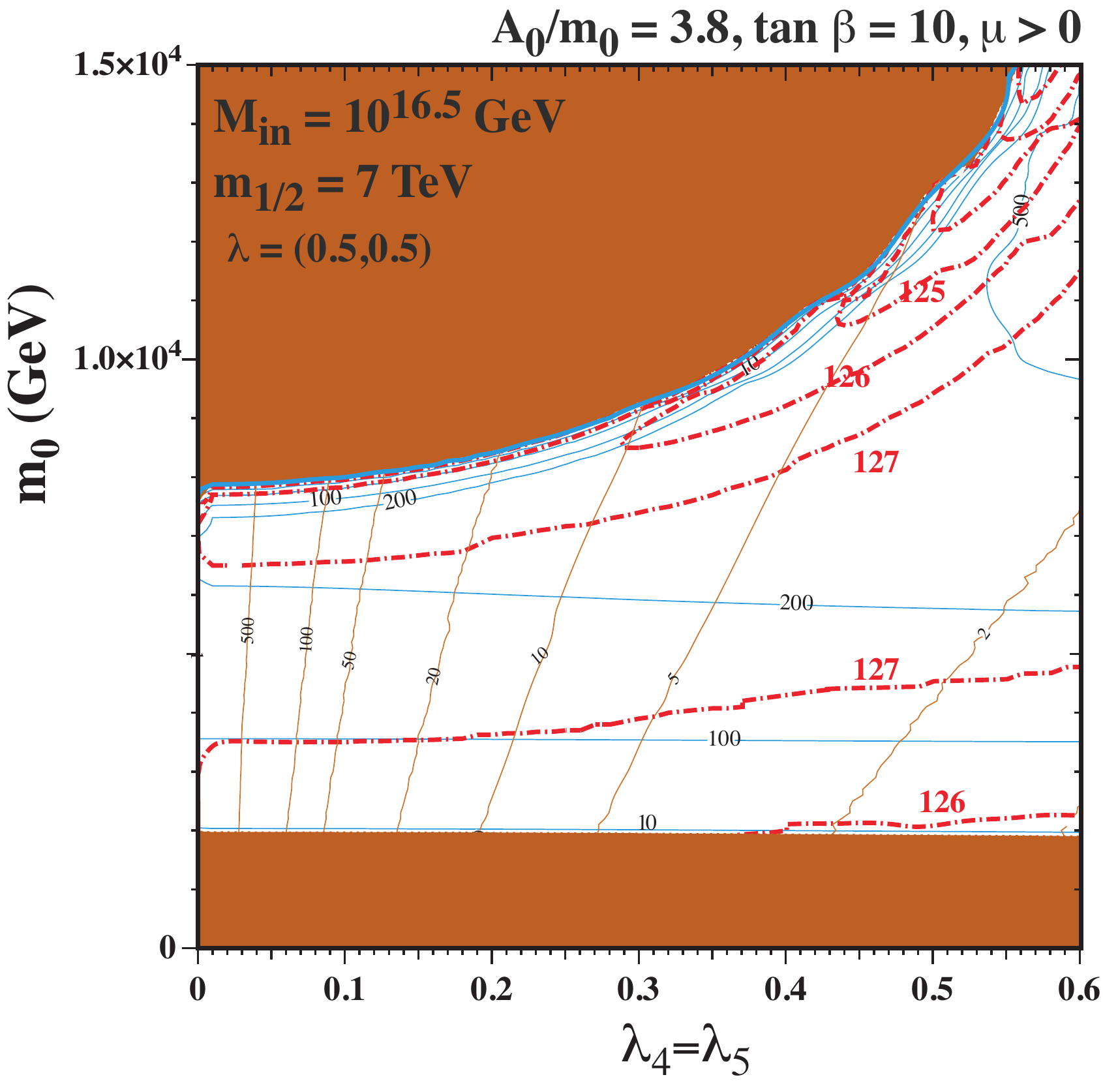}
\hspace{0.25in}
\includegraphics[width=7cm]{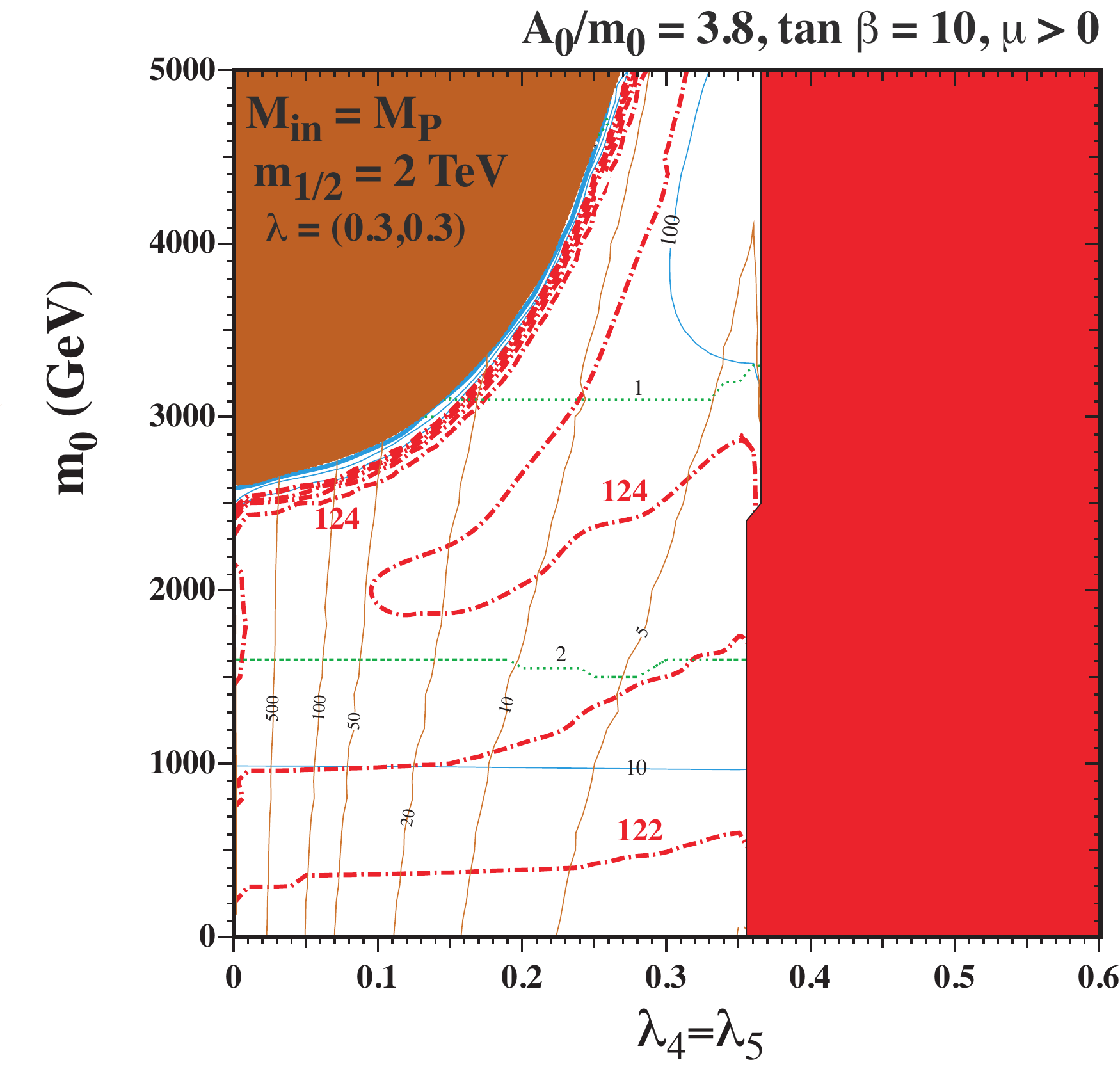}
\caption{\it Representative ($\lambda_4 = \lambda_5, m_0$) planes 
in the FSU(5) GUT model, both with $A_0/m_0 = 3.8$. 
In the left panel $M_{in} = 10^{16.5}$~GeV with $m_{1/2} = 7$~TeV, 
whereas in the right panel $M_{in} = M_P$ with $m_{1/2} = 2$~TeV. 
The brown shaded regions are excluded because the LSP is charged or
tachyonic, and the RGEs break down in the red shaded region at large
$\lambda_4 = \lambda_5$ in the right panel.
The  contours are as in Fig.~\ref{fig:plane}.}
\label{fig:lamvsm0A0}
\end{figure}

In Fig.~\ref{fig:lamvsm0A0}, we show a pair of ($\lambda_{4,5}, m_0$) planes with $A_0/m_0 = 3.8$ and 
$M_{in} = 10^{16.5}$ GeV (left panel) and $M_{in} = M_P$ (right panel),
as in Fig.~\ref{fig:stopplanes}. As previously, we see brown shaded
regions where the lightest neutralino is not the LSP, and a red shaded region
in the right panel where the RGEs break down.
We again see stop strips. For the lower value of $M_{in}$, we choose $m_{1/2} = 7$~TeV and for $M_{in} = M_P$, we take $m_{1/2} = 2$~TeV. We find that
$\Delta a_\mu$ is small everywhere in the left plane due to the large 
value of $m_{1/2}$, whereas in the right plane
we see contours of $\Delta a _\mu = 1$ and $2 \times 10^{-11}$, 
also too small to make a significant contribution to resolving the discrepancy
between experiment and the Standard Model calculation. 
As previously, we find that the proton lifetime is minimized,
and potentially observable, for large $\lambda_4 = \lambda_5$
and small $m_0$ when $M_{in} = 10^{16.5}$~GeV, 
whereas the proton lifetime is generally longer 
when $M_{in} = M_P$.

\subsection{Non-Universal Models and $\mathbf{g_\mu -2}$}
\label{sec:nonuniversal}

From the results in the previous subsection, it is clear that the contribution to $\Delta a_\mu$ is always small when  
universal boundary conditions applied for scalar and gaugino masses at $M_{in}$. Indeed, in all of the above planes, 
a significant contribution to $\Delta a_\mu$ occurs only at low supersymmetric masses that are in tension with LHC constraints and 
where the Higgs mass is well below the experimental value, even with a conservative
assessment of the theoretical uncertainty in the calculation of $m_h$. 
On the other hand, previous analyses have shown that
substantially larger contributions to $\Delta a_\mu$
are possible when some degree of universality is abandoned \cite{eenno1,king}.

Therefore, in this subsection, we depart from full gaugino and scalar mass universality at $M_{in}$, while retaining the constraints imposed by FSU(5). 
Thus, we include two independent gaugino masses, a common mass $M_5$ 
for the SU(5) gauginos $\tilde g, \tilde W$ and $\tilde B$, and an independent mass $M_{X1}$
for the `external' gaugino $\tilde B_X$. This is to be contrasted with our previous assumption that
$M_5 = M_{X1} = m_{1/2}$ at $M_{in}$. 
Similarly we now include five independent soft
supersymmetry-breaking scalar masses, $m_{10}$ for sfermions in the $\mathbf{10}$ 
representations of SU(5), $m_{\bar 5}$ for sfermions in the $\mathbf{\overline 5}$
representations of SU(5), $m_1$ for the right-handed sleptons in the singlet
representations, and two Higgs soft masses $m_{H_{u,d}}$ for the MSSM Higgs doublets stemming from $\mathbf{5}$ and $\mathbf{\overline 5}$ representations of SU(5).  Previously we had set $m_{10} = m_{\bar 5} = m_1 = m_{H_u} = m_{H_d} =  m_0$. 
Guided by the results of \cite{eenno1}, we make the illustrative choices $\tan \beta = 35$, $A_0/m_0 = 1.8$, $M_{in} = 10^{16.5}$ GeV, $M_5 = 2.4$ TeV, $m_{10} = 1$ TeV, $m_{H_d} = -4.72$ TeV, and $m_{H_u} = -5.1$ TeV. For the latter two,
the signs refer to the sign of mass-squared, and these choices correspond to the choice of $\mu = 4.77$ TeV and a pseudo scalar mass, $m_A = 2.1$ TeV for the example in \cite{eenno1}. These were found to optimize the value of $\Delta a_\mu$. 

\begin{figure}[!ht]
\centering
\includegraphics[width=7cm]{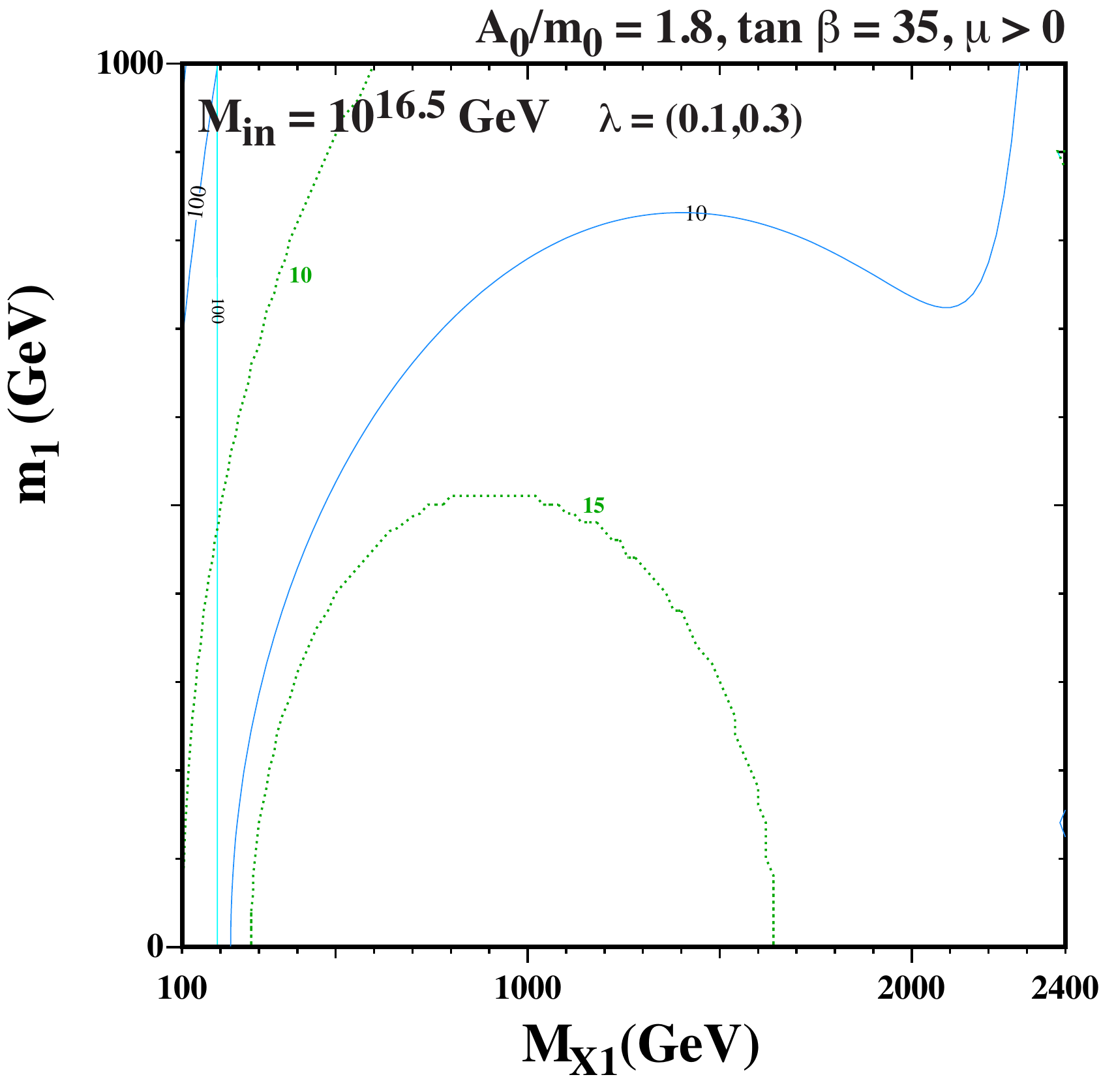}
\hspace{0.25in}
\includegraphics[width=7cm]{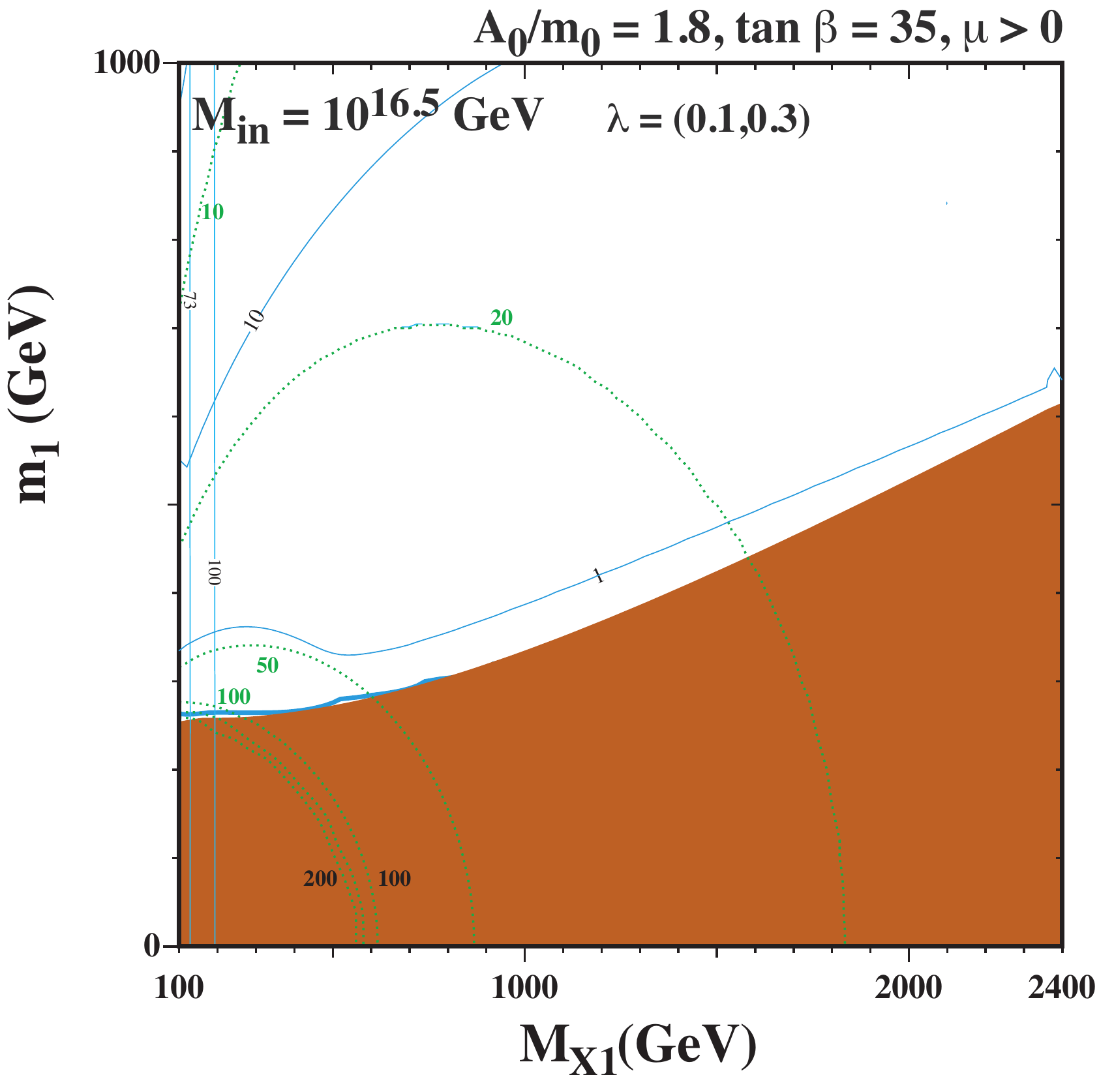}
\caption{\it Some $(M_{X1}, m_1)$ planes in the FSU(5) GUT model with $M_5 = 2.4$ TeV, $m_{10} = 1$ TeV, $\tan \beta = 35$, $A_0/m_{10} = 1.8$. We assume $m_{\bar 5} = m_{10}$ in the left panel and $m_{\bar 5} = m_{10}/2$ in the right panel. The  contours are as in Fig.~\ref{fig:plane}.}
\label{fig:m1M1}
\end{figure}

Some results are displayed in the $(M_{X1}, m_1)$ planes shown in Fig.~\ref{fig:m1M1}.
In the left panel, only the singlet masses $M_{X1}$ and $m_1$ break universality, i.e., we set
$m_{\bar 5} = m_{10} = 1$ TeV in this case. The Higgs mass varies very little in this plane and is always slightly larger than 122 GeV (no contours are shown). Similarly the proton lifetime varies very little and is approximately $(1.2 \pm 0.6)\times 10^{36}$ years.  In contrast, the relic density (indicated by the labeled blue contours) varies significantly, reaching values as large as $\Omega h^2 = 500$ in the upper left corner of the panel.
Also seen as vertical light blue lines
are the lower limits to the mass of the lightest gaugino 
$m_\chi > 100$~GeV (which is valid for generic slepton masses)
and $> 73$~GeV (which can be reached if the mass difference between the LSP
and the lightest slepton $< 2$~GeV).
Finally, we
show as green dotted lines some contours of $\Delta a_\mu$ in units of $10^{-11}$. In general, they are significantly larger than was found in the universal case, with contours of 10 and $15 \times 10^{-11}$ appearing, the largest value of $\Delta a_\mu$ being $18 \times 10^{-11}$.
While an improvement over the universal case, these are still too small 
to account for the discrepancy between the Standard Model and experiment. 

We can increase $\Delta a_\mu$
by choosing a lower value of $m_{\bar 5}$ relative to $m_{10}$.  
As an example, in the right panel of Fig.~\ref{fig:m1M1}
we take $m_{\bar 5} = m_{10}/2$. The lower right portion of the figure,
shaded brown, is excluded because the LSP is charged, namely
the right-handed selectron/smuon.
The $\Omega_\chi h^2 = 0.1$ contour appears along a
selectron/smuon coannihilation strip running close to the boundary
of the charged LSP region at small $m_1$. 
It is tracked by the $\Omega_\chi h^2 = 1$
contour at larger $m_1$.  
The Higgs mass values are similar to those in the left panel,
varying very slowly and always about 122.3 GeV across the plane. 
The proton lifetime is also nearly constant at around
$(1.1 \pm 0.6)\times 10^{36}$ years. However, the contribution to 
$a_\mu$ is now significantly larger, as there are contours of 10 
(in the upper left corner of the panel), 20, 50 and (in the lower left 
corner) 100, 150 and 200, 
again in units of $10^{-11}$. However, 
$\Delta a_\mu = 200 \times 10^{-11}$ appears outside the 
selectron/smuon LSP region only when the lightest gaugino mass
is below its lower limit of 73~GeV.
However, the $150 \times 10^{-11}$ contour extends to the right of the
vertical LSP mass limit of 100 GeV, in a region where the gaugino is the 
LSP and has a relic density $\Omega_\chi h^2 \sim 0.1$.
This region resembles the best $\Delta a_\mu$ point found in \cite{eenno1}. 

\begin{figure}[!ht]
\centering
\includegraphics[width=7cm]{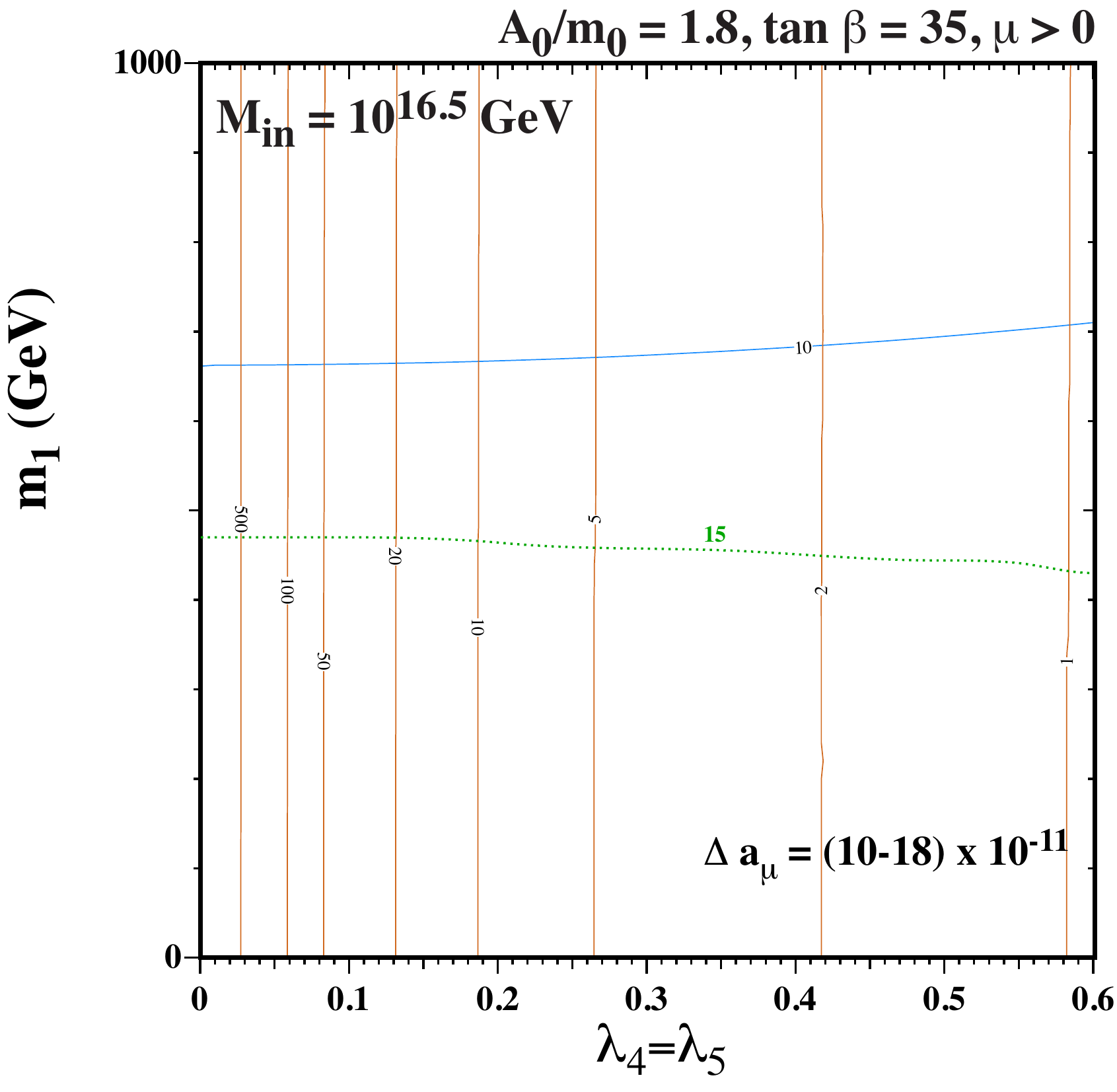}
\hspace{0.25in}
\includegraphics[width=7cm]{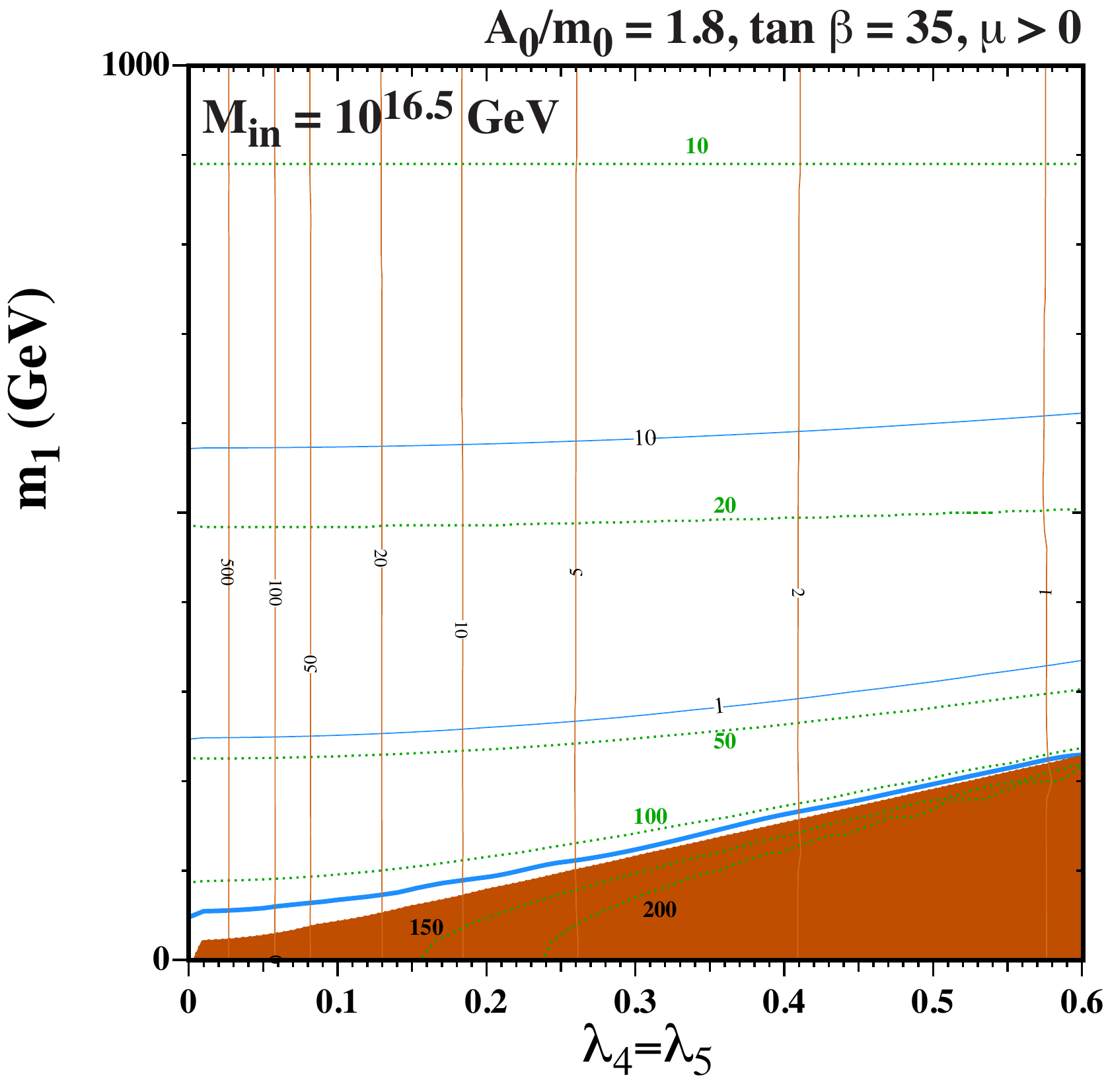}
\caption{\it Representative $(\lambda_4=\lambda_5, m_1)$ planes in the
FSU(5) GUT model. In both panels, $A_0/m_0 = 1.8$, $\tan \beta = 35$ and
$M_{in} = 10^{16.5}$ GeV. In the left panel $M_5 = 2.4$~TeV, 
$M_{X1} = 0.8$~TeV and $m_{10} = m_{\bar 5} = 1$~TeV, while
in the right panel $M_5 = 2.4$~TeV, $M_{X1} = 0.2$~TeV, $m_{10} = 1$~TeV
and $m_{\bar 5} = 0.5$~TeV. The  contours are as in Fig.~\ref{fig:plane}.}
\label{fig:lamvsm0NU}
\end{figure}

The sensitivity to $\lambda_{4,5}$ for similar choices
of model parameters is shown in the left panel of Fig.~\ref{fig:lamvsm0NU}, which displays a $(\lambda_{4,5}, m_1)$ plane with $M_{X1} = 0.8$~TeV
and other parameters the same as those used in the left panel of Fig.~\ref{fig:m1M1}. 
In this case, we see only a single relic density contour,
which has $\Omega_\chi h^2 = 10$.
The Higgs mass is again slightly larger than 122 GeV across the plane, 
and $\Delta a_\mu \simeq (10-18) \times 10^{-11}$. However, we now
see a large variation in the proton lifetime, which varies from 
$5 \times 10^{37}$ years at low values of $\lambda_4 = \lambda_5$, 
to $10^{35}$ years at large values. 

In contrast, in the right panel of Fig.~\ref{fig:lamvsm0NU}
we fix $M_{X1} = 200$~GeV, with $m_{10} = 2m_{\bar 5} = 1$~TeV. 
The proton lifetime decreases as $\lambda_4 = \lambda_5$ increase,
becoming potentially observable for values $\gtrsim 0.4$ (taking
into account the matrix element uncertainties).
The Higgs mass is not very sensitive to the choice of $m_{\bar 5}$ 
or the change in $M_{X1}$, with the Higgs mass being slightly
above 122~GeV across the plane displayed. The relic density is decreased
at low $m_1$, as the mass of the selectron is lower and there is 
now a long relic density selectron/smuon coannihilation strip where 
$\Omega h^2 = 0.1$ just above the brown shaded region where the LSP is
a selectron or smuon.  The value of $\Delta a_\mu$ 
is now larger as well, and we see contours of 
10, 20, 50, $100 \times 10^{-11}$ all lying above the selectron/smuon
LSP region.~\footnote{However, contours of $\Delta a_\mu = 150$ and 
$200 \times 10^{-11}$ appear inside that region.} We note the appearance of a
`quadrifecta' strip at large $\lambda_4 = \lambda_5$ close to the charged-LSP
boundary, where $\Delta a_\mu \sim 100 \times 10^{-11}$, $\tau(p \to e^+ \pi^0)$
is potentially detectable, $m_h$ is compatible with experiment and
$\Omega_\chi h^2 \sim 0.1$ (though the latter is not a requirement, as
mentioned previously).

\begin{figure}[!ht]
\centering
\includegraphics[width=6cm]{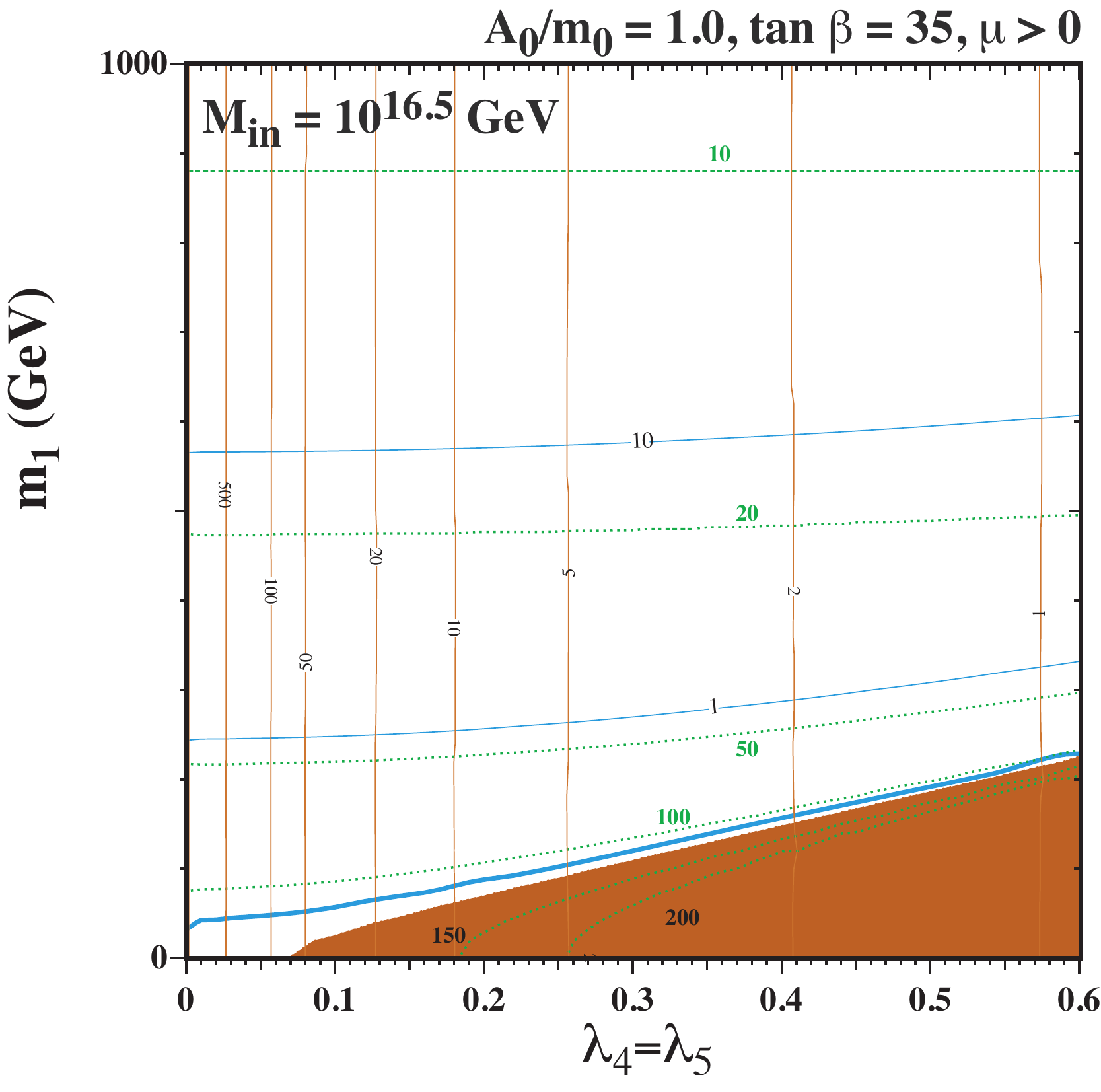}
\hspace{0.25in}
\includegraphics[width=6cm]{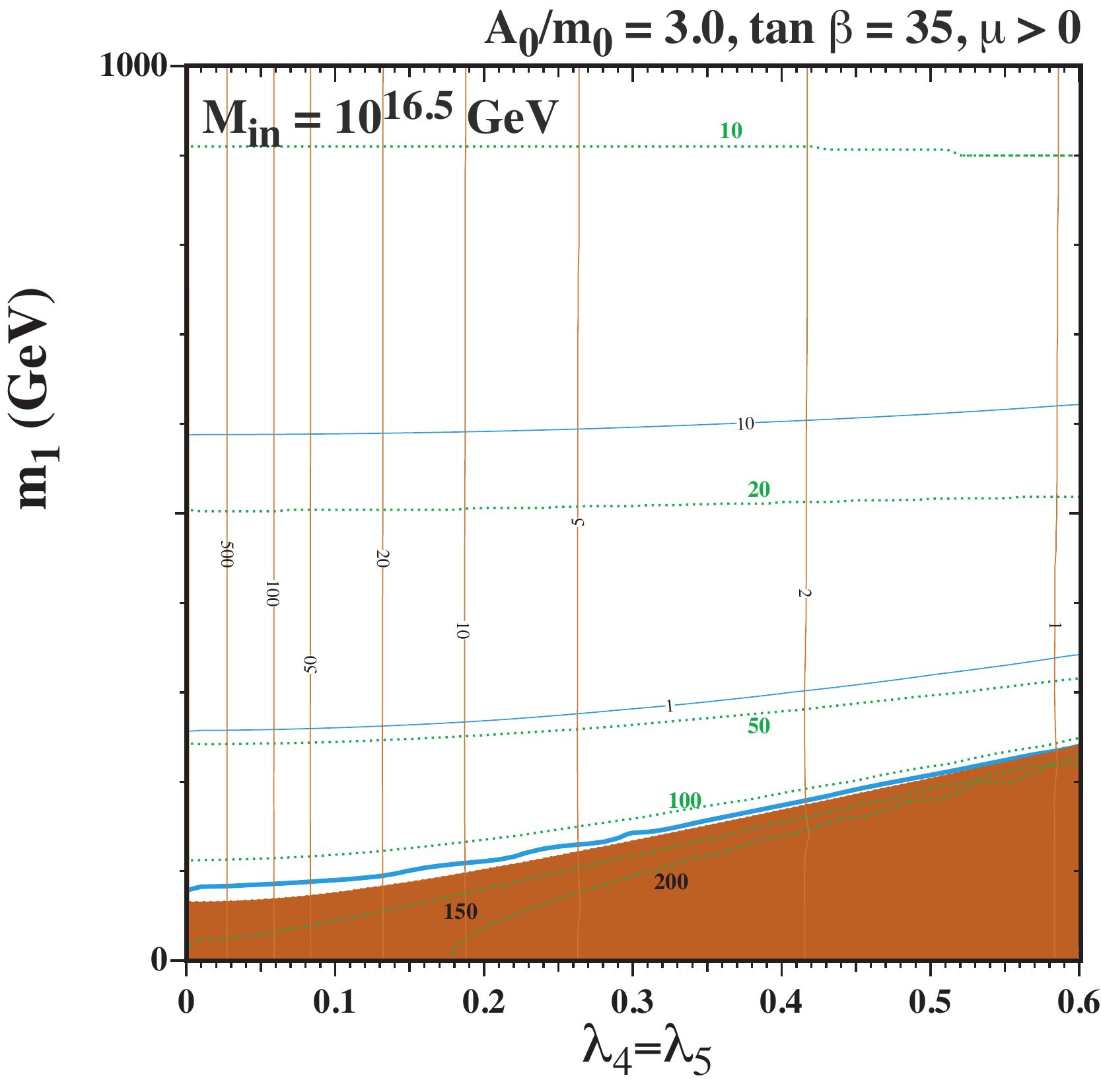}\\
\includegraphics[width=6cm]{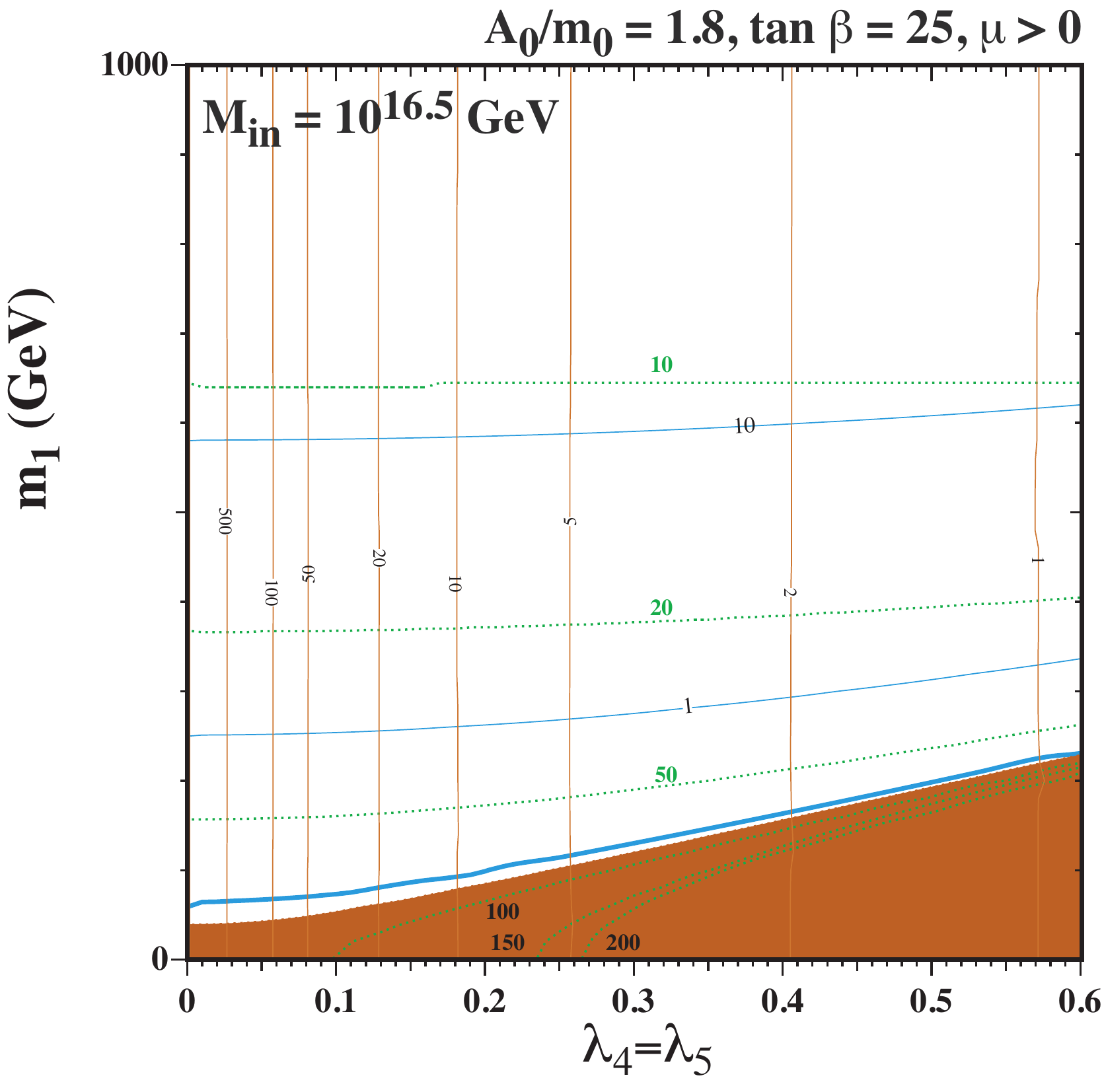}
\hskip 0.25in
\includegraphics[width=6cm]{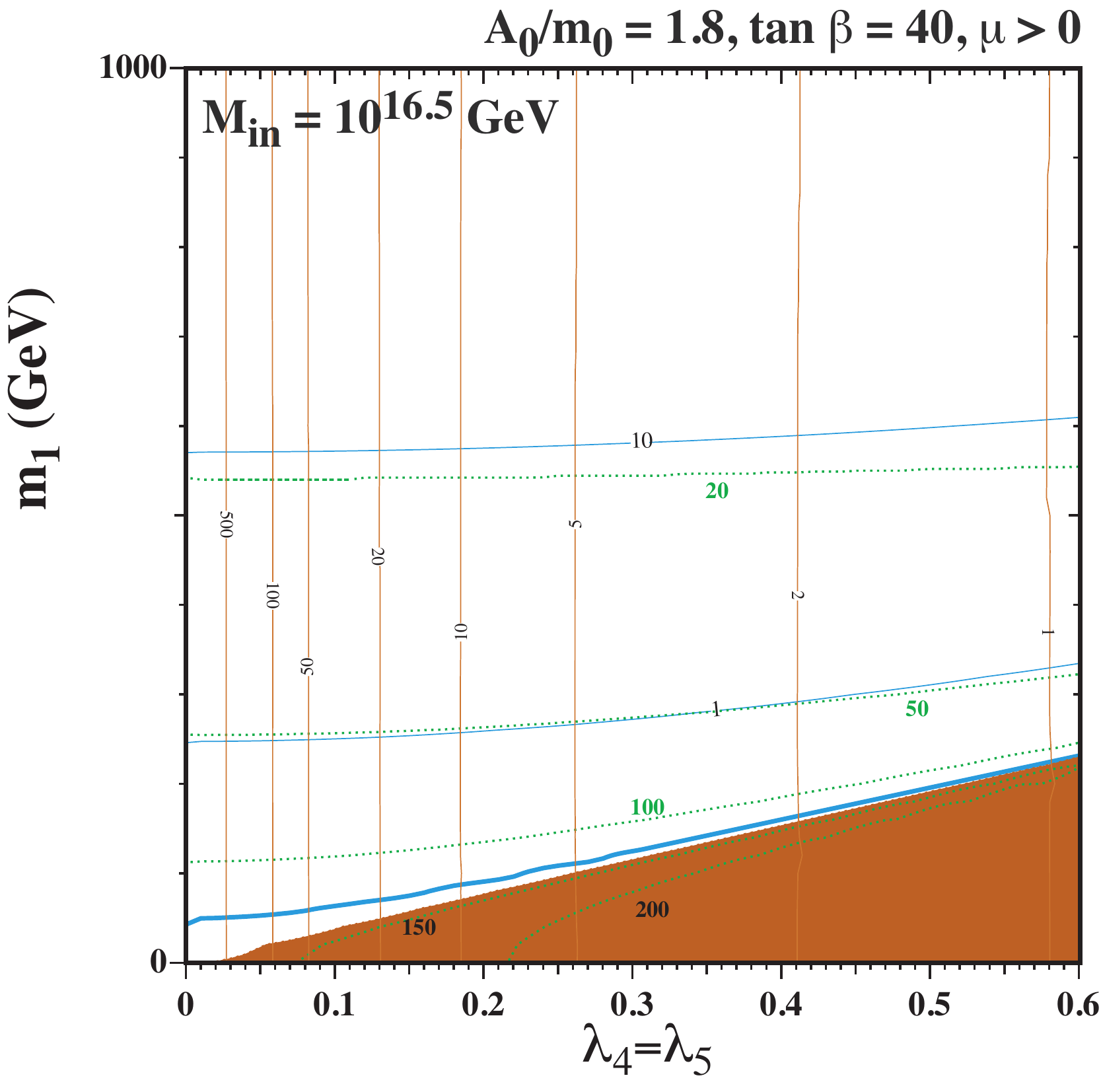}
\caption{\it As in Fig.~\ref{fig:lamvsm0NU}, but for different choices of $A_0/m_0 = 1, 3$
(upper left and upper right panels) and of $\tan \beta = 25, 40$ (lower left and right panels). The  contours are as in Fig.~\ref{fig:plane}.}
\label{fig:A0tbplanes}
\end{figure}

The sensitivities of the `quadrifecta' region to some of the input parameters
are shown in Fig.~\ref{fig:A0tbplanes}. In the upper pair of panels we explore
the sensitivity to $A_0/m_0$, which is taken to be 1 and 3 in the left and right
panels, respectively. We see that there is rather small sensitivity to $A_0/m_0$,
with $\Delta a_\mu$ and the proton lifetime both increasing slightly with the
value of $A_0/m_0$. In the lower pair of panels we explore the sensitivity to $\tan \beta$,
which is taken to be 25 and 40 in the left and right panels, respectively.
We see that $\Delta a_\mu$ is significantly smaller for $\tan \beta = 25$ and larger
for $\tan \beta = 40$, whereas the proton lifetime is quite insensitive to $\tan \beta$.
We recall that the validity of the perturbation regime is quite restricted for
larger values of $\tan \beta$, and recall that our results are rather insensitive to
the value of $\lambda_6$.

\section{Conclusions}
\label{sec:conx}

We have considered in this paper various aspects of the phenomenology of
supersymmetric flipped SU(5) GUTs, focusing on predictions for proton decay and 
$g_\mu - 2$. We have found that, if the soft supersymmetry-breaking
parameters are constrained to be universal at some high scale
$M_{in}$ above the GUT scale, the proton lifetime is typically
$\gtrsim 10^{36}$~yrs and the supersymmetric contribution to $g_\mu - 2$ is small.
The proton lifetime is generally too long to be detected in the foreseeable
future, and the model does not contribute significantly to reducing the tension
between data-driven calculations of $g_\mu -2$ within the
Standard Model and the experimental measurement. However, we have found that there is
a region of the constrained flipped SU(5) parameter space with large 
$\mathbf{ 10 \ 10 \ 5}$ and $\mathbf{ \overline{10} \ \overline{10} \ \overline{5}}$
couplings where $p \to e^+ \pi^0$ decay may be detectable in the Hyper-Kamiokande 
experiment~\cite{HK} now under construction. Nevertheless, the flipped SU(5) GUT
contribution to $g_\mu -2$ is still small. 

However, we have found that if the 
universality constraints on the soft supersymmetry-breaking masses are relaxed
there is a region of flipped SU(5) GUT parameter space where the model
contribution to $g_\mu -2$ can be large enough to reduce significantly the
discrepancy between theory and experiment while $\tau (p \to e^+ \pi^0)$ may
simultaneously be short enough to be detected in Hyper-Kamiokande. This region appears 
when $A_0/m_0 \sim 1 - 3$ and $\tan \beta \sim 25 - 40$, for suitable values 
of the other flipped SU(5) parameters. We call this the `quadrifecta' region,
since the theoretical calculation of the light Higgs mass is compatible
with the experimental measurement, within uncertainties, and the strip where
the relic LSP density $\Omega_\chi h^2 \simeq 0.12$ if the Universe expands
adiabatically passes through the region.~\footnote{However, as we have
emphasized in the text, analyses of flipped SU(5) 
cosmology~\cite{egnno2,egnno3,egnno4,FSU5Cosmo} suggest that
there may have been significant entropy generation during the expansion of the Universe,
in which case this value of $\Omega_\chi h^2$ is not required.}

This `quadrifecta' region was previously identified in a dedicated analysis of
$g_\mu -2$ in the flipped SU(5) GUT~\cite{eenno1}, and it is encouraging that
this region appears quite stable under mild variations in the input parameters.
As pointed out in~\cite{eenno1}, in this region both the LSP and lighter smuon masses are
very close to the LEP lower limits on their masses of $\sim 100$~GeV, and
detection of the LSP, smuon and selectron should be possible at the LHC. Their
discovery would be a striking success for the flipped SU(5) framework
described here, which could be complemented by the detection of $p \to e^+ \pi^0$ 
decay in the Hyper-Kamiokande experiment~\cite{HK}.

\section*{Acknowledgements}

The work of J.E. was supported partly by the United Kingdom STFC Grant ST/T000759/1 
and partly by the Estonian Research Council via a Mobilitas Pluss grant. 
The work of N.N. was supported by the Grant-in-Aid for Scientific Research B (No.20H01897), 
Young Scientists (No.21K13916), and Innovative Areas (No.18H05542).
The work of D.V.N. was supported partly by the DOE grant DE-FG02-13ER42020 and partly by the Alexander S. Onassis Public Benefit Foundation.
The work of K.A.O. was supported partly
by the DOE grant DE-SC0011842 at the University of Minnesota.

\end{document}